\documentclass[journal,onecolumn,12pt]{IEEEtran}
\usepackage{amssymb}
\usepackage{amsmath}
\usepackage{graphicx}
\usepackage{tcolorbox}
\usepackage{mathtools}
\usepackage{braket}

\usepackage[ruled]{algorithm2e}
\usepackage{bm}
\usepackage{romannum}
\usepackage{tabularx,ragged2e,booktabs,caption}
\usepackage{xcolor}
\usepackage{enumitem}
\usepackage{amsmath}
\usepackage{hyperref}
\hypersetup{
    colorlinks=true,
    linkcolor=black,
    filecolor=black,      
    urlcolor=black,
}
\makeatletter
\newcommand{\norm}[1]{\left\lVert#1\right\rVert}
\newcommand{\E}{\operatorname*{\mathbb{E}}\ilimits@}
\makeatother

\DeclareMathOperator*{\argmin}{arg\,min}

\begin{document}
\title{Predicting Multi-Antenna Frequency-Selective Channels via Meta-Learned Linear Filters based on Long-Short Term Channel Decomposition}
\author{Sangwoo Park,~\IEEEmembership{Member,~IEEE} and
        Osvaldo Simeone,~\IEEEmembership{Fellow,~IEEE}

\thanks{This work was presented in part at IEEE ICASSP 2022 \cite{park2021predicting}.} 
\thanks{S. Park and O. Simeone are with the Department of Engineering, King's College London, London WC2R 2LS, U.K. (e-mails: \{sangwoo.park, osvaldo.simeone\}@kcl.ac.uk).}
\thanks{The work of O. Simeone was supported by the European Research Council (ERC) under the European Union's Horizon 2020 research and innovation programme (grant agreement No. 725731).}
}
\maketitle

\begin{abstract}
An efficient data-driven prediction strategy for multi-antenna frequency-selective channels must operate based on a small number of pilot symbols. This paper proposes novel channel prediction algorithms that address this goal by integrating transfer and meta-learning with a reduced-rank parametrization of the channel. The proposed methods optimize linear predictors by utilizing data from previous frames, which are generally characterized by distinct propagation characteristics, in order to enable fast training on the time slots of the current frame. The proposed predictors rely on a novel long-short-term decomposition (LSTD) of the  linear prediction model that leverages the disaggregation of the channel into long-term space-time signatures and fading amplitudes. We first develop predictors for single-antenna frequency-flat channels based on transfer/meta-learned quadratic regularization. Then, we introduce  transfer and meta-learning algorithms for LSTD-based prediction models that build on equilibrium propagation (EP) and alternating least squares (ALS). Numerical results under the 3GPP 5G standard channel model demonstrate the impact of transfer and meta-learning on reducing the number of pilots for channel prediction, as well as the merits of the proposed LSTD parametrization.
\end{abstract}
\IEEEpeerreviewmaketitle

\section{Introduction}
\label{sec:intro}
The capacity to accurately predict channel state information (CSI) is a key enabler of proactive resource allocation strategies, which are central to many visions for efficient and low-latency communications in 6G and beyond (see, e.g., \cite{tang2019future}). The problem of channel prediction is relatively straightforward in the presence of known channel statistics. In fact, under the common assumption that multi-antenna frequency-selective channels follow stationary complex Gaussian processes,   optimal channel predictors can be obtained via linear minimum mean squared error (LMMSE) estimators such as the Wiener filter \cite{hoeher1997two}. However, in practice, the channel statistics are not known, and predictors need to be optimized based on training data obtained through the transmission of pilot signals \cite{baddour2005autoregressive, duel2000long, liu2013mimo,min2007mimo, komninakis2002multi, kashyap2017performance}. The problem addressed by this paper concerns the design of data-efficient channel predictors for multi-antenna frequency-selective channels.

\begin{figure}
    \centering
    \hspace{-0cm}
    \includegraphics[width=0.6\columnwidth]{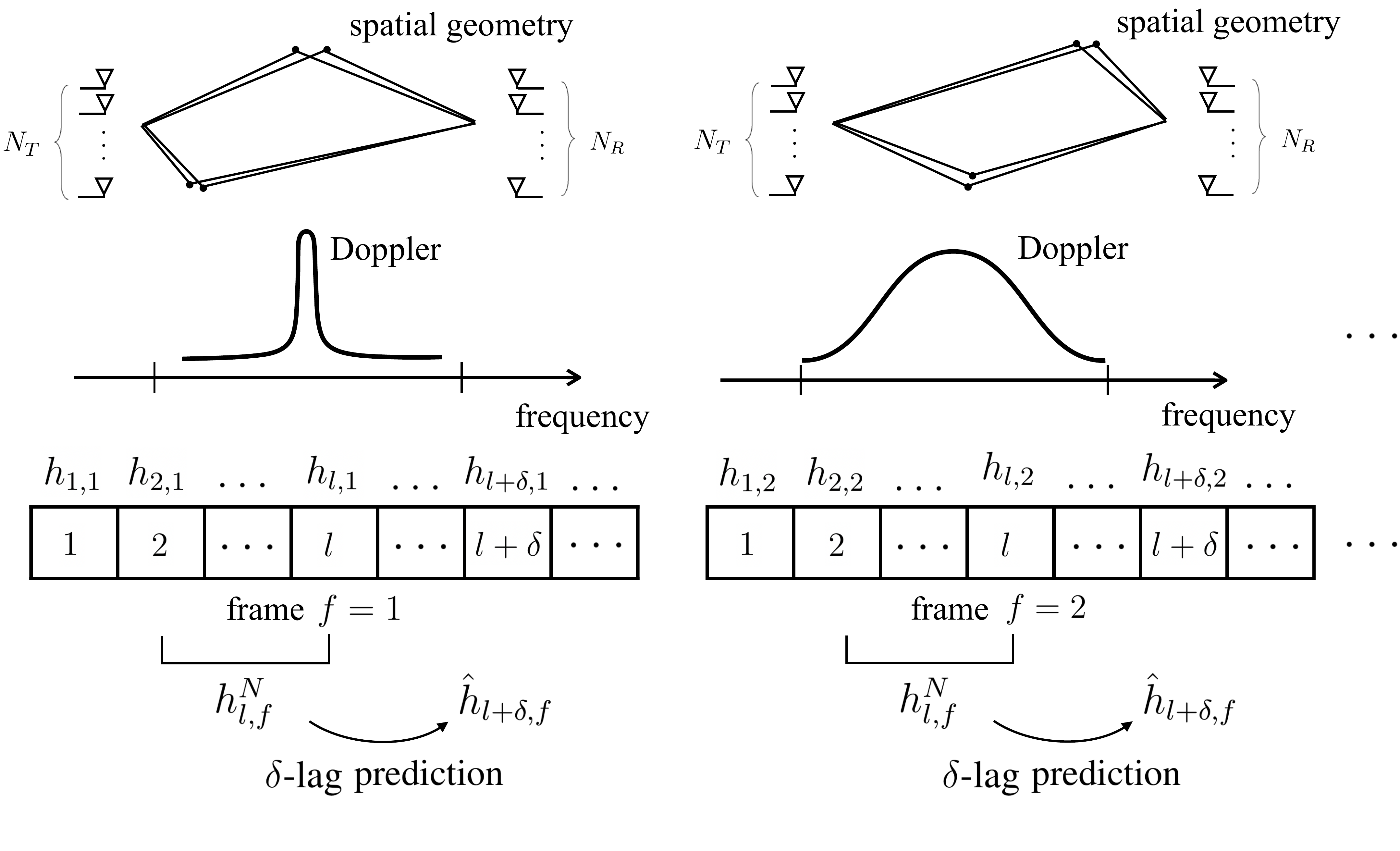}
    \caption{Illustration of the frame-based transmission system under study: At any frame $f$, based on the previous $N$ channels $h_{l,f}^N$, we investigate the problem of optimizing the $\delta$-lag prediction $\hat{h}_{l+\delta,f}.$}
    \label{fig:system_model}
    \vspace{-0.5cm}
\end{figure}

\subsection{Context and Prior Art}
A classical approach to tackle this problem is to optimize linear predictors obtained via autoregressive (AR) models \cite{baddour2005autoregressive, duel2000long, liu2013mimo} or Kalman filtering (KF) \cite{min2007mimo, komninakis2002multi, kashyap2017performance} by estimating channel statistics from the available pilot data. More recently, deep learning-based nonlinear predictors have also been proposed to adapt to channel statistics through training of neural networks,  namely recurrent neural network \cite{liu2006recurrent, jiang2019comparison, jiang2020long, kibugi2021machine}, convolutional neural network \cite{yuan2020machine, zhang2022predicting}, and multi-layer perceptrons \cite{kim2020massive}.  

As reported in \cite{jiang2019comparison, jiang2020long, kibugi2021machine, kim2020massive}, deep learning based predictors tend to require larger training (pilot) data, while failing to outperform well-designed linear filters in the low-data regime. Some solutions addressing this issues include  \cite{bogale2020adaptive}, which applies reinforcement learning to determine whether to predict channels or not at current time; and the use of hypernetworks to adapt parameters of a KF accordingly to current channel dynamics \cite{pratik2021neural}.


Most prior work, with the notable exception of \cite{ pratik2021neural}, focuses on the optimization of channel predictors under the assumption of a stationary spatio-temporal correlation function across the time interval of interest. This conventional approach fails to leverage common structure that may exist across multiple \emph{frames}, with each frame being characterized by distinct spatio-temporal correlations (see Fig. 1). Reference \cite{ pratik2021neural} allowed for varying Doppler spectra across frames, through a deep learning-based hypernetwork that is used to adapt the parameters of a generative model \cite{pratik2021neural}. 

This paper takes a different approach that allows us to move beyond the single-antenna setting studied in \cite{pratik2021neural}. As described in the next subsection, key ingredients of the proposed methods are transfer and meta-learning. Transfer learning \cite{torrey2010transfer} and meta-learning \cite{thrun2012learning} aim at using knowledge from distinct tasks in order to reduce the data requirements on a new task of interest. Given a large enough resemblance between different tasks, both transfer learning and meta-learning have shown remarkable performance to reduce the sample complexity in general machine learning problems \cite{raghu2019rapid}. Transfer learning applies to a specific target task, while meta-learning caters for adaptation to any new task (see, e.g., \cite{jose2021information}). 

 Previous applications of transfer learning to communication systems include beamforming for multi-user multiple-input single-output (MISO) downlink   \cite{yuan2020transfer} and for intelligent reflecting surfaces (IRS)-assisted MISO downlink \cite{ge2021beamforming}; and  downlink channel prediction \cite{yang2020deep,parera2019transfer} (see also  \cite{yuan2020transfer, yang2020deep}). Meta-learning has been applied to communication system, including to demodulation \cite{park2020learning,cohen2021learning,mao2019roemnet,raviv2021meta}; decoding  \cite{jiang2019mind}; end-to-end design of encoding and decoding with and without a channel model \cite{park2020meta, park2020end}; MIMO detection \cite{goutay2020deep}; beamforming for multiuser MISO downlink systems via  \cite{zhang2021embedding}; layered division multiplexing for ultra-reliable
communications \cite{karasik2021learning}; UAV trajectory design  \cite{hu2021distributed}; and resource allocation  \cite{nikoloska2021black}.




\subsection{Contributions}

This paper proposes novel efficient data-driven channel prediction algorithms that reduce pilot requirements by integrating transfer and meta-learning with a novel long-short-term decomposition (LSTD) of the linear predictors. Unlike the prior art reviewed above, the proposed methods apply to multi-antenna frequency-selective channels whose statistics change across frames (see Fig. 1). Specific contributions are as follows.\\
\noindent $\bullet$ We develop efficient predictors for single-antenna frequency-flat channels based on transfer/meta-learned quadratic regularization. Transfer and meta-learning are used to leverage data from multiple frames in order to extract shared useful knowledge that can be used for prediction on the current frame (see Fig. 2).\\
\noindent $\bullet$  Targeting multi-antenna frequency-selective channels, we introduce the LSTD-based model class of linear predictors that builds on the well-known disaggregation of standard channel models into long-term space-time signatures and fading amplitudes \cite{liu2013mimo, simeone2004lower, cicerone2006channel, pedersen2000stochastic, abdi2002space}. Accordingly, the channel is described by multipath features, such as angle of arrivals, delays, and path loss,  that change slowly across the frame, as well as by fast-varying fading amplitudes.  Transfer and meta-learning algorithms for LSTD-based prediction models are proposed that build on equilibrium propagation (EP) and alternating least squares (ALS).\\
\noindent $\bullet$  Numerical results under the 3GPP 5G standard channel model demonstrate the impact of transfer and meta-learning on reducing the number of pilots for channel prediction, as well as the merits of the proposed LSTD parametrization.

Part of this paper was presented in \cite{park2021predicting}, which only covered meta-learning for the case of single-antenna frequency-flat channels. As compared to  \cite{park2021predicting}, this journal version includes both transfer and meta-learning, and it addresses the general scenario of multi-antenna frequency-selective channels by introducing and developing the LSTD model class of linear predictors.

\subsection{Organization}

The rest of the paper is organized as follows. In Sec.~\ref{sec:system_model}, we detail system and channel model, and describe conventional, transfer, and meta-learning concepts. In Sec.~\ref{sec:single_angenna}, we develop solutions for single-antenna frequency-flat channels. In Sec.~\ref{sec:multi_antenna}, multi-antenna frequency-selective channels are considered, and we propose  LSTD-based linear prediction schemes. Numerical results are presented in Sec.~\ref{sec:experiments}, and conclusions are presented in Sec.~\ref{sec:conclusion}. 

\textit{Notation:} In this paper, $(\cdot)^\top $ denotes the transposition; $(\cdot)^\dagger$ the Hermitian transposition; $(\cdot)_F$ the Frobenius norm; $|\cdot|$ the absolute value; $||\cdot||$ the Euclidean norm; $\text{vec}(\cdot)$ the vectorization operator that stacks the columns of a matrix into a column vector; $[\cdot]_{i}$ the $i$-th element of the vector; and {\color{black}$I_S$ the  $S \times S$ identity matrix for some integer $S$.}
\section{System Model}
\label{sec:system_model}
\subsection{System Model}
\label{subsec:basic_system_model}

As shown in Fig.~\ref{fig:system_model}, we study a frame-based transmission system, with each frame containing multiple time slots. Each frame carries data from a possibly different user to the same receiver, e.g., a base station. The receiver has $N_R$ antennas, while the transmitters have $N_T$ antennas. The channel $h_{l,f}$ in slot $l=1,2,\ldots$ of frame $f=1,2,\ldots$ is a vector with $S=N_RN_TW$ entries, with $W$ being the delay spread measured in number of transmission symbols within each frame $f$, the multi-path channels $h_{l,f} \in \mathcal{C}^{N_R N_T W\times 1}$ are characterized by fixed, frame-dependent, average path powers, path delays, Doppler spectra, and angles of arrival and departure \cite{3gpp_tr_901}. For instance, in a frame $f$, we may have a slow-moving user in line-of-sight condition subject to time-invariant fading, while, in another, the channel may have significant scattering with fast temporal variations with a large Doppler frequency. In both cases, the frame is assumed to be short enough that average path powers, path delays, Doppler spectra, and angles of arrival and departure do not change within the frame \cite{simeone2004lower,cicerone2006channel}.

As also seen in Fig.~\ref{fig:system_model}, for each frame $f$, we are interested in addressing the lag-$\delta$ channel prediction problem, in which channel $h_{l+\delta,f}$ is predicted based on the $N$ past channels  
\begin{align}
H^N_{l,f} =  [h_{l,f}, \ldots, h_{l-N+1,f}] \in \mathcal{C}^{S \times N}.
\label{eq:H_lf^N_original}
\end{align}
We adopt linear prediction with regressor $V_f \in \mathcal{C}^{SN \times S}$, so that the prediction is given as 
\begin{align}
\hat{h}_{l+\delta,f} = V_f^\dagger \text{vec}(H^N_{l,f}).
\label{eq:predicted_channel}
\end{align}
The focus on linear prediction is justified by the optimality of linear estimation for Gaussian stationary processes \cite{gallager2008principles}, which provide standard models for fading channels in rich scattering environments.

Assuming no prior knowledge of the channel model, we adopt a data-driven approach to the design of the predictor \eqref{eq:predicted_channel}. Accordingly, to train the linear predictor \eqref{eq:predicted_channel}, for any frame $f$, the receiver is assumed to have available the training set
\begin{align}
\mathcal{Z}^\text{tr}_f = \{ (x_{i,f}, y_{i,f})\}_{i=1}^{L^\text{tr}} \equiv \{(\text{vec}(H_{l,f}^N),  h_{l+\delta,f})\}_{l=N}^{L^\text{tr}+N-1} 
\label{eq:training_dataset}
\end{align}
encompassing $L^\text{tr}$ input-output examples. Data set $\mathcal{Z}_f^\text{tr}$ can be constructed from $L^\text{tr}+N+\delta-1$ channels $\{ h_{1,f}, \ldots, h_{L^\text{tr}+N+\delta-1,f} \}$ by using the lag-$\delta$ channel $h_{l+\delta,f}$ as label for the covariate vector $\text{vec}(H_{l,f}^N)$. In practice, the channel vectors $h_{l,f}$ are estimated using pilot symbols, and estimation noise can be easily incorporated in the model {\color{black}(see Sec.~\ref{subsec:noisy})}. Throughout, we implicitly assume that the channels $h_{l,f}$ correspond to estimates available at the receiver. 

From data set $\mathcal{Z}_f^\text{tr}$ in \eqref{eq:training_dataset}, we write the corresponding $L^\text{tr} \times SN$ input matrix $X_f^\text{tr} = [x_{1,f}^\dagger,\ldots,x_{L^\text{tr},f}^\dagger]^\top$, and the $L^\text{tr} \times S$ target matrix $Y_f^\text{tr}=[y_{1,f}^\dagger,\ldots,y_{L^\text{tr},f}^\dagger]^\top$, so that the data set can be expressed as the pair $\mathcal{Z}_f^\text{tr} = (X_f^\text{tr}, Y_f^\text{tr})$.

\subsection{Channel Model}
\label{subsec:channel_model}
We adopt the standard spatial channel model \cite{3gpp_tr_901}. Accordingly, a channel vector $h_{l,f}$ for slot $l$ in frame $f$, is obtained by sampling the continuous-time multipath vector channel impulse response
\begin{align}
    h_{l,f}(\tau) = \sum_{d=1}^{D} \sqrt{\Omega_{d,f}}a_{d,f} g(\tau - \tau_{d,f})\text{exp}(-j\color{black}{2\pi \gamma_{d,f}}t_l),
    \label{eq:multivariate_channel_model}
\end{align}
which is the sum of contributions from $D$ paths. In \eqref{eq:multivariate_channel_model}, the waveform $g(\tau)$ is given by the convolution of the transmitted waveform and the matched filter at the receiver. Furthermore, the contribution of the $d$-th path depends on the average power $\Omega_{d,f}$; the path delay $\tau_{d,f}$; the $N_TN_R \times 1$ spatial vector $a_{d,f}$; the {\color{black} Doppler frequency $\gamma_{d,f}$}; and the starting wall-clock time of the $l$-th slot $t_l$. The average power $\Omega_{d,f}$, path delays $\tau_{d,f}$, spatial vector $a_{d,f}$, and Doppler frequency $\gamma_{d,f}$ are constant within one frame since they depend on large-scale geometrical features of the propagation environment. However, they may change over frames following Clause 7.6.3.2 (Procedure B) in \cite{3gpp_tr_901}. The number of paths is assumed without loss of generalization to be the same for all frames $f$ since one can set $\Omega_{d,f}=0$ for frames with a smaller number of paths.


In \cite{3gpp_tr_901}, the spatial vector $a_{d,f}$ has a structure that depends on field patterns and steering vectors of the transmit and receive antennas, as well on the polarization of the antennas. Mathematically, the entry of the spatial vector $a_{d,f}$ corresponding to the receive and transmit antenna element $n_R$ and $n_T$ can be modeled as \cite{3gpp_tr_901}
\begin{align}
[a_{d,f}]_{n_R+(n_T-1)N_R} = \mathbf{F}_{ rx,n_R}(\theta_{d,f,ZOA}, \phi_{d,f,AOA})^T  \mathbf{M}_{d,f}   \mathbf{F}_{tx,n_T}( \theta_{d,f,ZOD},\phi_{d,f,AOD})   \exp{\left(-\frac{j2\pi l_{d,f,n_R,n_T}}{\lambda_0}\right)},
\label{eq:3gpp_channel_model}
\end{align}
where $\mathbf{F}_{rx,n_R}(\cdot,\cdot)$ and $\mathbf{F}_{tx,n_T}(\cdot,\cdot)$ are the $2 \times 1$ field patterns; $\theta_{d,f,ZOA}$, $\phi_{d,f,AOA}$,  $\theta_{d,f,ZOD}$, $\phi_{d,f,AOD}$ are the zenith angle of arrival (ZOA), azimuth angle of arrival (AOA), zenith angle of departure (ZOD), azimuth angle of departure (AOD) (in degrees); $\lambda_0$ is the wavelength (in $\text{m}$) of the carrier frequency; $l_{d,f,n_R,n_T}$ is the length of the path (in $\text{m}$) between the two antennas; and $\mathbf{M}_{d,f}$ is the polarization coupling matrix defined as
\begin{align}
    \mathbf{M}_{d,f} = \left(\begin{array}{ll} \quad \quad \exp{\left(j\Phi^{\theta \theta}_{d,f}\right)} \quad\quad \sqrt{1/\kappa_{d,f}} \exp{\left( j \Phi_{{d,f}}^{\theta \phi} \right)} \\ \sqrt{1/\kappa_{{d,f}}} \exp{\left( j \Phi_{{d,f}}^{\phi \theta} \right)}\quad\quad \exp{\left(j\Phi^{\phi \phi}_{d,f}\right)} \end{array} \right),
\end{align}
with random initial phase $\Phi_{d,f}^{(\cdot, \cdot)} \sim U(-\pi,\pi)$ and log-normal distributed cross polarization power ratio (XPR) $\kappa_{d,f}>0$ \cite{3gpp_tr_901}. 

In order to obtain the $S\times 1$ vector $h_{l,f}$, we sample the continuous-time channel $h_{l,f}(\tau)$ in \eqref{eq:multivariate_channel_model} at Nyquist rate $1/T$ to obtain $W$ discrete-time $N_R N_T \times 1$ channel impulse response  \begin{align}
    h_{l,f}[w] = h_{l,f}((w-1)T)
    \label{eq:final_channel_vector_tap}
\end{align} for $w=1,\ldots,W$. Following \cite{simeone2004lower}, the channel vector $h_{l,f}\in\mathcal{C}^{N_R N_T W \times 1}$ is obtained by concatenating the $W$ channel vectors $h_{l,f}[w]$ for $w=1,\ldots, W$ as
\begin{align}
    h_{l,f} = [h_{l,f}[1], \ldots, h_{l,f}[W]]^\top.
    \label{eq:final_channel_vector}
\end{align}


\subsection{Conventional Learning}
The optimization of the linear predictor $V_f$ in \eqref{eq:predicted_channel} can be formulated as a supervised learning problem as it will be detailed in Sec.~\ref{sec:single_angenna}. In conventional learning, the predictor $V_f$ is designed separately in each frame $f$ based on the corresponding data set $\mathcal{Z}_f^\text{tr}$. 
In order for this predictor $V_f$ to generalize well to slots in the same frame $f$ outside the training set, it is necessary to have a sufficiently large number of training slots, $L^\text{tr}$ \cite{simeone2018brief}. 

\subsection{Transfer and Meta-Learning}
\begin{figure}
    \centering
    \hspace{-0cm}
    \includegraphics[width=1\columnwidth]{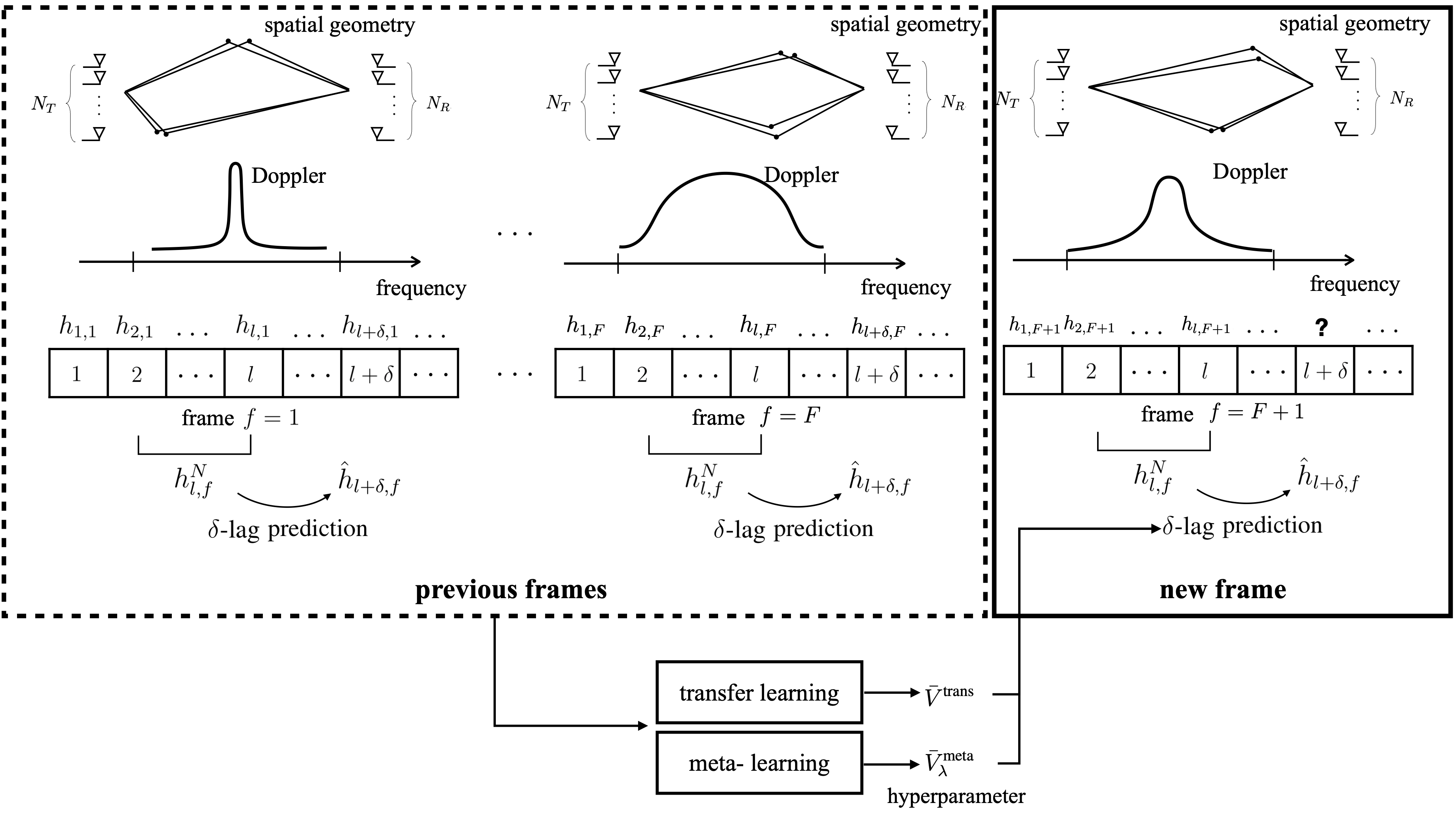}
    \caption{Illustration of the considered transfer and meta-learning methods: With access to pilots from previously received frames, transfer and meta-learning aim at obtaining the hyperparameters $\bar{V}$ to be used for channel prediction in a new frame.}
    \label{fig:transfer_and_meta_sys}
    \vspace{-0.5cm}
\end{figure}
In conventional learning, the number of required training slots $L^\text{tr}$ can be reduced by selecting hyperparameters in the learning problem that reflect prior knowledge about the prediction problem at hand. In the next sections, we will explore solutions that optimize such hyperparameters based on data received from multiple previous frames. To this end, as illustrated in Fig.~\ref{fig:transfer_and_meta_sys}, we assume the availability of channel data collected from $F$ frames received in the past. In each frame, the channel follows the model described in Sec.~\ref{subsec:channel_model}. Accordingly, data from previous frames consists of $L+N+\delta-1$ channels $\{ h_{1,f}, \ldots, h_{L+N+\delta-1,f} \}$ for some integer $L$. 

Using these channels, the data set 
\begin{align}
    \mathcal{Z}_f  = \{(x_{i,f}, y_{i,f})\}_{i=1}^{L} \equiv \{(\text{vec}(H_{l,f}^N), h_{l+\delta,f})\}_{l=N}^{L+N-1}
    \label{eq:Z_f_def}
\end{align}
can be obtained as explained in Sec.~\ref{subsec:basic_system_model}, where $L$ is typically larger than $L^\text{tr}$, although this will not be assumed in the analysis.
Correspondingly, we also define the $L \times N$ input matrix $X_f$ and the $L \times 1$ target vector $y_f$. We will propose methods that leverage the historical knowledge available from data set $\mathcal{Z}_f$ for $f=1,\ldots,F$ via transfer and meta-learning with the goal of reducing number of pilots, $L^\text{tr}$, needed for channel prediction in a new frame (i.e., frame $F+1$ in Fig.~\ref{fig:transfer_and_meta_sys}).


{\color{black}
\subsection{Incorporating Estimation Noise}
\label{subsec:noisy}
Until now, we assumed that channel vectors $h_{l,f}$ are available noiselessly to the predictor. In practice, channel information needs to be estimated via pilots. To elaborate on this point let us assume the received signal model
\begin{align}
    y_{l,f}[i] = h_{l,f} x_{l,f}[i] + n_{l,f}[i],
    \label{eq:channel_model_pilot}
\end{align}
where $x_{l,f}[i]$ stands for the $i$th transmitted pilot symbol in block $l$ of frame $f$, $y_{l,f}[i]$ for the corresponding received signal, $h_{l,f}$ for the channel with additive white complex Gaussian noise $n_{l,f}[i] \sim \mathcal{CN}(0,N_0I_S)$. Given an average energy constraint $\mathbb{E}[x_{l,f}[i]^2]=E_x$ for the training symbol, the average Signal-to-Noise Ratio (SNR) is given as $E_x/N_0$. From \eqref{eq:channel_model_pilot}, we can estimate the channel as
\begin{align}
\check{h}_{l,f} = \frac{y_{l,f}[i]}{x_{l,f}[i]} = h_{l,f} + \frac{n_{l,f}[i]}{x_{l,f}[i]} = h_{l,f} + \xi,
\end{align}
which suffers from channel estimation noise $\xi \sim \mathcal{CN}(0, \text{SNR}^{-1}I_S)$. If $P$ training symbols are available in each block, the channel estimation noise can be reduced via averaging to $\text{SNR}^{-1}/P$. Channels $\check{h}_{l,f}$ can be used as training data in the schemes described in the previous subsections.

}
\section{Single-Antenna Frequency-Flat Channels}
\label{sec:single_angenna}
In this section, we propose transfer learning and meta-learning methods for single-antenna flat-fading channels, which result in $S=1$. Throughout this section, we write the prediction matrix $V_f \in \mathcal{C}^{SN \times S}$ in \eqref{eq:predicted_channel} as the vector $v_f \in \mathcal{C}^{N \times 1}$, and the target data $Y_f^\text{tr} \in \mathcal{C}^{L^\text{tr} \times S}$ as the vector $y_f^\text{tr} \in \mathcal{C}^{L^\text{tr} \times 1}$. Correspondingly, we rewrite the linear predictor \eqref{eq:predicted_channel} as
\begin{align}
\hat{h}_{l+\delta,f} = v_f^\dagger \text{vec}(H^N_{l,f}).
\label{eq:predicted_channel_scalar}
\end{align}

\subsection{Conventional Learning}
Assuming the standard quadratic loss, we formulate the supervised learning problem as the ridge regression optimization
\begin{align}
v^*(\mathcal{Z}_f^\text{tr} | \bar{v}) 
= \argmin_{v_f \in \mathcal{C}^{N \times 1}} \Big\{ || X_f^\text{tr} v_f - y_f^\text{tr} ||^2 + \lambda || v_f - \bar{v} ||^2 \Big\},
\label{eq:general_ridge_scalar}
\end{align}
with hyperparameters $(\lambda, \bar{v})$ given by the scalar $\lambda > 0$ and by the $N \times 1$ bias vector $\bar{v}$. The bias vector $\bar{v}$ can be thought of defining the prior mean of the predictor $v_f$, while $\lambda >0$ specifies the precision (i.e., inverse of the variance) of this prior knowledge. The solution of problem \eqref{eq:general_ridge_scalar} can be obtained explicitly as
\begin{align}
v^*(\mathcal{Z}_f^\text{tr} | \bar{v}) = ({A_f^\text{tr}})^{-1} \big( (X_f^\text{tr})^\dagger y_f^\text{tr} + \lambda \bar{v} \big),  \textrm{ with } A_f^\text{tr} = (X_f^\text{tr})^\dagger X_f^\text{tr} + \lambda I.
\label{eq:general_ridge_sol_scalar}
\end{align}


\subsection{Transfer Learning}
\label{sec:scalar_transfer}
Transfer learning uses data sets $\mathcal{Z}_f$ in \eqref{eq:Z_f_def} from the previous $F$ frames, i.e., with $f=1,\ldots,F$, to optimize the hyperparameter vector $\bar{v}$ in \eqref{eq:general_ridge_scalar} as
\begin{align}
    \bar{v}^{\text{trans}} &= \argmin_{v\in\mathcal{C}^{N\times 1}} \left\{ \sum_{f=1}^F \norm{X_f v - y_f}^2 \right\}.
\label{eq:transfer_objective}
\end{align}
The rationale for this choice is that vector $\bar{v}^\text{trans}$ provides a useful prior mean to be used in the ridge regression problem \eqref{eq:general_ridge_scalar}, since it corresponds to an optimized predictor for the previous frames. Having optimized the bias vector $\bar{v}^\text{trans}$,  we train a channel predictor $v$ via ridge regression \eqref{eq:general_ridge_scalar} using the training data $\mathcal{Z}_{f^\text{new}}$ for a new frame $f^\text{new}$ with $L^\text{new}$ training samples, to obtain
\begin{align}
v_{f^\textrm{new}}^* = v^*(\mathcal{Z}_{f^\text{new}}|\bar{v}^{\text{trans}}).
\label{eq:mte_scalar_transfer}
\end{align}


\subsection{Meta-Learning}
\label{sec:scalar_meta}
Unlike transfer learning, which utilizes all the available data sets $\{\mathcal{Z}_f\}_{f=1}^F$ from the previous frames at once as in \eqref{eq:transfer_objective}, meta-learning allows for the separate adaptation of the predictor in each frame. To this end, for each frame $f$, we split the $L$ data points into $L^\textrm{tr}$ training pairs $\{(x_{i,f}, y_{i,f})\}_{i=1}^{L^\textrm{tr}} \equiv \{(x_{i,f}^\textrm{tr}, y_{i,f}^\textrm{tr})\}_{i=1}^{L^\textrm{tr}}=\mathcal{Z}_f^\text{tr}$ and $L^\textrm{te}=L-L^\textrm{tr}$ test pairs $ \{(x_{i,f}, y_{i,f})\}_{i=L^\textrm{tr}+1}^{L} \equiv \{(x_{i,f}^\textrm{te}, y_{i,f}^\textrm{te})\}_{i=1}^{L^\textrm{te}}=\mathcal{Z}_f^\text{te}$, resulting in two separate data sets $\mathcal{Z}_f^\text{tr}$ and $\mathcal{Z}_f^\text{te}$. We correspondingly define the $L^\text{tr} \times N$ input matrix $X_f^\text{tr}$ and the $L^\text{tr} \times 1$ target vector $y_f^\text{tr}$, as well as the $L^\text{te} \times N$ input matrix $X_f^\text{te}$ and the $L^\text{te} \times 1$ target vector $y_f^\text{te}$.

The hyperparameter vector $\bar{v}$ is then optimized by minimizing the sum-loss of the predictors $v^{*}(\mathcal{Z}^\text{tr}_f|\bar{v})$ in \eqref{eq:general_ridge_scalar} that are adapted separately for each frame $f=1,\ldots,F$ given the bias vector $\bar{v}$. Accordingly, estimating the loss in each frame $f$ via the test set $\mathcal{Z}^\text{te}_f$ yields the meta-learning problem
\begin{align}
\bar{v}^{\text{meta}} &= \argmin_{\bar{v}\in\mathcal{C}^{N\times 1}} \left\{ \sum_{f=1}^F \big|v^*(\mathcal{Z}_f^\textrm{tr}|\bar{v})^\dagger x_{i,f}^\textrm{te} - y_{i,f}^\textrm{te}\big|^2 \right\}
\label{eq:meta_objective}
\end{align}

As studied in \cite{denevi2018learning}, the minimization in \eqref{eq:meta_objective} is a least squares problem that can be solved in closed form as
\begin{align}
    \bar{v}^{\text{meta}} &= \argmin_{\bar{v} \in\mathcal{C}^{N\times 1}} \sum_{f=1}^F \norm{\tilde{X}_f^\text{te}\bar{v} - \tilde{y}_f^\text{te}}^2 \nonumber\\
&= (\tilde{X}^\dagger \tilde{X})^{-1} \tilde{X}^\dagger \tilde{y},\label{eq:closed_from_sol_meta_scalar}
\end{align}
where $L^\text{te} \times N$ matrix $\tilde{X}_f^\text{te}$ contains by row the Hermitian transpose of the $N \times 1$ pre-conditioned input vectors $\{ \lambda (A_f^\text{tr})^{-1} x_{i,f}^\text{te} \}_{i=1}^{L^\text{te}}$, with $A_f^\text{tr} = (X_f^\text{tr})^\dagger X_f^\text{tr} + \lambda I$,; $\tilde{y}_f^\text{te}$ is $L^\text{te} \times 1$ vector containing vertically the complex conjugate of the transformed outputs 
$\{ (y^\text{te}_{i,f} - (y_f^\text{tr})^\dagger X_f^\text{tr} (A_{f}^\text{tr})^{-1} x_{i,f}^\text{te} \}_{i=1}^{L^\text{te}}$;
the $F L^\text{te} \times N$ matrix $\tilde{X} = [\tilde{X}_1^\text{te}, \ldots,\tilde{X}_F^\text{te}]^\top$ stacks vertically the $L^\text{te} \times N$ matrices $\{\tilde{X}_f^\text{te}\}_{f=1}^F$; and the $F L^\text{te} \times 1$ vector $\tilde{y} = [\tilde{y}_{1}^\text{te},\ldots,\tilde{y}_{F}^\text{te}]^\top$ stacks vertically the $L^\text{te}\times 1$ vectors $\{\tilde{y}_f^\text{te}\}_{f=1}^F$. 

After meta-learning, similar to transfer learning, based on the meta-learned hyperparameter $\bar{v}^{\text{meta}}_\lambda$, we train a channel predictor via ridge regression \eqref{eq:general_ridge_scalar}, obtaining 
\begin{align}
v_{f^\textrm{new}}^* = v^*(\mathcal{Z}_{f^\text{new}}|\bar{v}^{\text{meta}}).
\label{eq:mte_scalar}
\end{align}
\section{Multi-Antenna Frequency-Selective Channels}
\label{sec:multi_antenna}
In this section, we study the more general scenario with any number of antennas and with frequency-selective channels, resulting in $S > 1$. As we will discuss, a na\"ive extension of the techniques presented in the previous sections is undesirable, since this would not leverage the structure of the channel model \eqref{eq:multivariate_channel_model}. For this reason, in the following, we will introduce novel hybrid model- and data-driven solutions that build on the channel model \eqref{eq:multivariate_channel_model}.

\subsection{Na\"ive Extension}
We start by briefly presenting the direct extension of the approaches studied in the previous section to any $S > 1$. Unlike the previous section, we adopt the general matrix notation introduced in Sec.~\ref{sec:system_model}. First, with $S=1$, conventional learning obtains the predictor by solving problem  \eqref{eq:general_ridge_scalar}, which is generalized to any $S>1$ as the minimization
\begin{align}
V^*(\mathcal{Z}_f^\text{tr} | \bar{V}) = \argmin_{V_f \in \mathcal{C}^{SN \times S}} \Big\{ || X_f^\text{tr} V_f - Y_f^\text{tr} ||_F^2 + \lambda || V_f - \bar{V} ||_F^2 \Big\}
\label{eq:general_ridge_vec}
\end{align}
over the linear prediction matrix $V_f$ in \eqref{eq:predicted_channel}. Similarly, transfer learning computes the bias matrix $\bar{V}^\text{trans}$ by solving the following generalization of problem \eqref{eq:transfer_objective}
\begin{align}
    \bar{V}^{\text{trans}} &= \argmin_{V \in \mathcal{C}^{SN \times S}}\left\{ \sum_{f=1}^F \norm{X_f V - Y_f}_F^2 \right\}
    \label{eq:transfer_obj_vec_naive}
\end{align}
followed by the evaluation of the predictor $V^*(\mathcal{Z}_f^\text{tr}|\bar{V}^\text{trans})$ using \eqref{eq:general_ridge_vec};
while meta-learning addresses the following generalization of minimization \eqref{eq:meta_objective}
\begin{align}
\bar{V}^{\text{meta}} &= \argmin_{\bar{V} \in \mathcal{C}^{SN \times S}} \left\{ \sum_{f=1}^F \sum_{i=1}^{L^\textrm{te}}\big|V^*(\mathcal{Z}_f^\textrm{tr}|\bar{V})^\dagger x_{i,f}^\textrm{te} - y_{i,f}^\textrm{te}\big|^2 \right\}
\label{eq:meta_obj_vec_naive}
\end{align}
over the bias matrix $\bar{V} \in \mathcal{C}^{SN \times S}$, which is used to compute the predictor $V^*(\mathcal{Z}_f^\text{tr}|\bar{V}^\text{meta})$ in \eqref{eq:general_ridge_vec}.

The issue with the na\"ive extensions \eqref{eq:transfer_obj_vec_naive} and \eqref{eq:meta_obj_vec_naive} is that the dimension of the predictor $V$ and of the hyperparameter matrix $\bar{V}$ can become extremely large when $S$ grows. This, in turn, may lead to overfitting in the hyperparameter space \cite{yin2019meta} when the number of frames, $F$, is limited. This form of overfitting may prevent transfer learning and meta-learning from effectively reducing the sample complexity for problem \eqref{eq:general_ridge_vec}, since the optimized hyperparameter matrix $\bar{V}$ would be excessively dependent on the data received in the $F$ previous frames. To solve this problem, we propose next to utilize the structure of the channel model \eqref{eq:multivariate_channel_model} in order to reduce the dimension of the channel parametrization.

\subsection{Long-Short-Term Decomposition (LSTD) Channel Model}
\label{sec:subsec_LSTD}
The channel model \eqref{eq:multivariate_channel_model} implies that the channel vector $h_{l,f}$ in \eqref{eq:final_channel_vector_tap}--\eqref{eq:final_channel_vector}  can be written as the product of a frame-dependent $N_R N_T W \times D$ matrix $T_f$ and of a slot-dependent $D\times 1$ vector $\beta_{l,f}$ as \cite{simeone2004lower}
\begin{align}
    h_{l,f} = T_{f} \beta_{l,f},
    \label{eq:origianl_decomp}
\end{align}
where $T_f$ collects space-time signatures of the $D$ paths as
\begin{align}
    T_{f} = [\Omega_{1,f}^{1/2} \mathbf{g}(\tau_{1,f})\otimes \text{vec}(a_{1,f}), \ldots,  \Omega_{D,f}^{1/2} \mathbf{g}(\tau_{D,f})\otimes \text{vec}(a_{D,f} ) ],
    \label{eq:T_f}
\end{align}
with $\mathbf{g}(\tau_{d,f}) = [g(-\tau_{d,f}), \ldots, g((W-1)T - \tau_{d,f})]^\top$ being the $W \times 1$ vector that collects the Nyquist-rate samples of the delayed waveform $g(\tau-\tau_{d,f})$; and the $D \times 1$ fading amplitude vector being defined as $\beta_{l,f} = [\text{exp}(-jw_{1,f}t_l), \ldots, \text{exp}(-jw_{D,f}t_l)]^\top$.

The frame-dependent matrix $T_f$ is typically rank-deficient, since paths are generally not all resolvable \cite{swindlehurst1998time, nicoli2003multislot}. To account for this structural property of the channel, as in \cite{simeone2004lower}, we introduce a $N_RN_TW \times K$ full rank unitary matrix $B_f$ such that $\text{span}\{T_f\} = \text{span}\{B_f\}$ and redefine \eqref{eq:origianl_decomp} as
\begin{align}
    h_{l,f} = B_f d_{l,f}.
    \label{eq:new_decomposition}
\end{align}
As an example, the unitary matrix $B_f$ can be obtained from the singular value decomposition of matrix $T_f$, i.e., $T_f = B_f \Lambda^{1/2}_fU_f^\dagger$, by introducing the $K\times 1$  vector $d_{l,f} = \Lambda^{1/2}_f U_f^\dagger \beta_{l,f}$ \cite{simeone2004lower}. For future reference, we also rewrite \eqref{eq:new_decomposition} as
\begin{align}
    h_{l,f} = \sum_{k=1}^K b_f^k d_{l,f}^k,
    \label{eq:channel_parametrization}
\end{align}
where $d_{l,f}^k$ is the $k$-th element of the vector $d_{l,f}$ and $b_{f}^k$ is the $k$-th column of the matrix $B_f$. 

We will refer to matrix $B_f$ in \eqref{eq:channel_parametrization} as the \emph{long-term space-time feature matrix}, or \emph{feature matrix} for short, while vector $d_{l,f}$ will be referred as the \emph{short-term} corresponding \emph{amplitude vector}. parametrization \eqref{eq:new_decomposition}, or \eqref{eq:channel_parametrization}, is particularly efficient when the feature matrix $B_f$ can be accurately estimated from the available data. For conventional learning, this requires observing a sufficiently large number of slots per frame, i.e., a large $L^\text{new}$ \cite{simeone2004lower}, as well as a channel that varies sufficiently quickly across each frame. In contrast, as we will explore, transfer and meta-learning can potentially leverage data from multiple frames in order to enhance the estimation of the feature matrix.

\subsection{Long-Short-Term Decomposition (LSTD)-based Prediction Model}
\begin{figure}
    \centering
    \hspace{-0cm}
    \includegraphics[width=0.8\columnwidth]{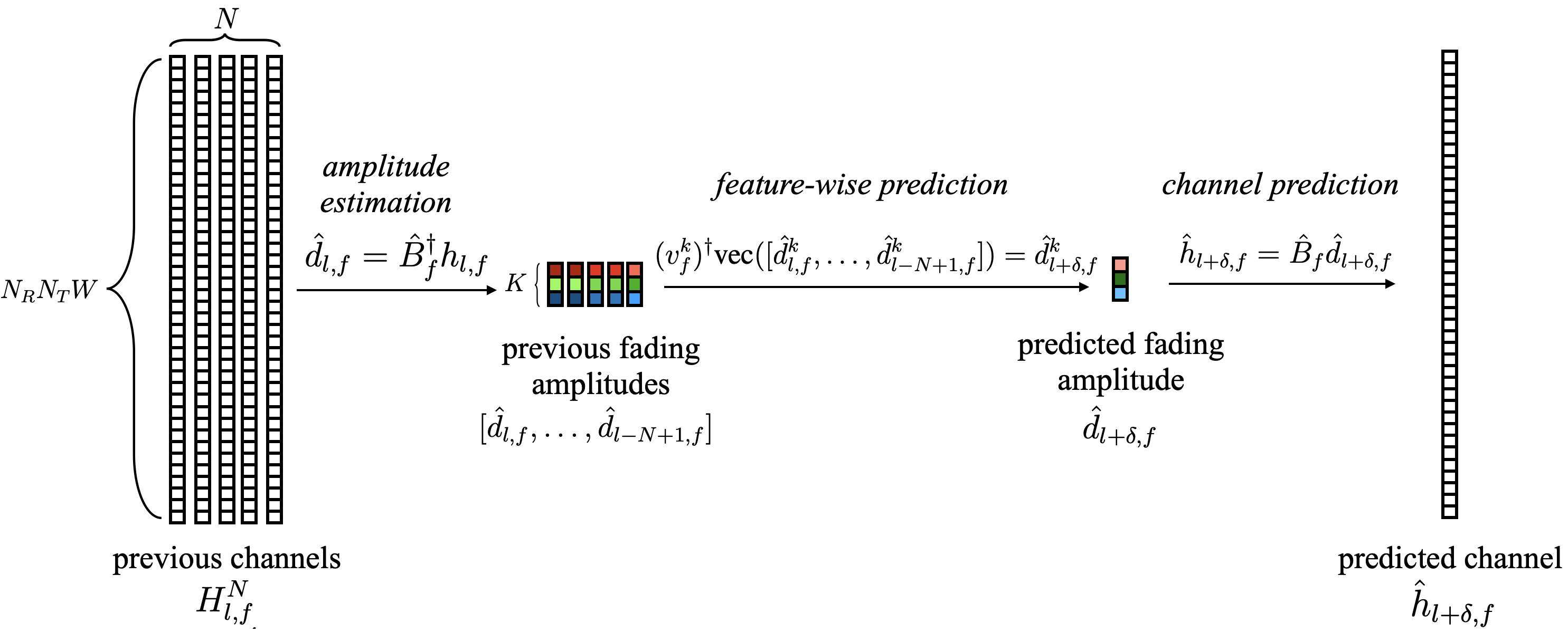}
    \caption{Illustration of the considered LSTD-based prediction model: \emph{(i)} estimate amplitudes $d_{l,f}$ via the estimated long-term feature matrix $\hat{B}_f$; \emph{(ii)} feature-wise short-term fading amplitude prediction $\hat{d}_{l+\delta,f}^k$ based on feature-wise predictor $v_{f}^k$ for $k=1,\ldots,K$; \emph{(iii)} reconstruction of the predicted channel $\hat{h}_{l+\delta,f}$ based on the feature matrix $\hat{B}_f$ and predicted fading amplitude $\hat{d}_{l+\delta,f}$. }
    \label{fig:reduced_rank_conven}
    \vspace{-0.5cm}
\end{figure}
Given the LSTD channel model \eqref{eq:new_decomposition}--\eqref{eq:channel_parametrization}, in this subsection we redefine the problem of predicting channel $h_{l+\delta,f}=B_f d_{l+\delta,f}$ as the problem of estimating the feature matrix $B_f$ and predicting the amplitude vector $d_{l+\delta,f}$ based on the available data. This will lead to a reduced-rank parametrization of the linear predictor \eqref{eq:predicted_channel}. 

To start, we write the predicted channel $\hat{h}_{l+\delta,f}$ as
\begin{align}
    \hat{h}_{l+\delta,f} = \hat{B}_f \hat{d}_{l+\delta,f},
    \label{eq:est_recon}
\end{align}
where $\hat{B}_f$ and $\hat{d}_{l+\delta,f}$ are the estimated feature matrix and the predicted amplitude vector, respectively. To define the corresponding predictor, we first observe that the input matrix $H_{l,f}^N$ in \eqref{eq:H_lf^N_original} can be expressed using \eqref{eq:new_decomposition} as
\begin{align}
    H_{l,f}^N = B_f[d_{l,f},\ldots,d_{l-N+1,f}].
    \label{eq:H_lf^N}
\end{align} 
Assume now that we have an estimated feature matrix $\hat{B}_f$. If this estimate is sufficiently accurate, the $N$ past amplitudes $[{d}_{l,f}, \ldots, {d}_{l-N+1,f}] \in \mathcal{C}^{K \times N}$ can be in turn estimated from $H_{l,f}^N$ as
\begin{align}
   [\hat{d}_{l,f}, \ldots, \hat{d}_{l-N+1,f}]  = \hat{B}_f^\dagger H_{l,f}^N.
   \label{eq:estimated_vector_amplitudes}
\end{align}
Consider now the prediction of the $k$-th amplitude $d_{l+\delta,f}^k$. Generalizing \eqref{eq:predicted_channel_scalar}, we adopt the linear predictor 
\begin{align}
    \hat{d}_{l+\delta,f}^k = ({v}_f^k)^\dagger \text{vec}([\hat{d}_{l,f}^k, \ldots, \hat{d}_{l-N+1,f}^k] ),
    \label{eq:pred_amplitude}
\end{align}
where $v_f^k$ is an $N \times 1$ prediction vector, and 
\begin{align}
    [\hat{d}_{l,f}^k, \ldots, \hat{d}_{l-N+1,f}^k] = (\hat{b}_f^k)^\dagger H_{l,f}^N \in \mathcal{C}^{1 \times N}
    \label{eq:past_amplitude}
\end{align}
is the $k$-th row of the matrix \eqref{eq:estimated_vector_amplitudes}, which represents the past $N$ fading scalar amplitudes that correspond to the $k$-th feature $b_f^k$. Plugging the prediction \eqref{eq:pred_amplitude} into \eqref{eq:est_recon} yields the predicted channel $\hat{h}_{l+\delta,f}$ (cf. \eqref{eq:channel_parametrization})
\begin{align}
    \hat{h}_{l+\delta,f} = \sum_{k=1}^K \hat{b}_f^k \hat{d}_{l+\delta,f}^k.
    \label{eq:pred_channel_est_vec}
\end{align}

As detailed in Appendix~\ref{app:vf_rr}, inserting \eqref{eq:pred_amplitude} and \eqref{eq:past_amplitude} to \eqref{eq:pred_channel_est_vec}, we can express the LSTD-based prediction \eqref{eq:pred_channel_est_vec} in the form \eqref{eq:predicted_channel} as
\begin{align}
\hat{h}_{l+\delta,f} = (V_f^{(K)})^\dagger \text{vec}(H^N_{l,f}),
\label{eq:predicted_channel_reduced_rank}
\end{align}
where the LSTD-based predictor matrix $V_f^{(K)} \in \mathcal{C}^{SN \times S}$ is given as 
\begin{align}
    V_f^{(K)} = \sum_{k=1}^K {v}_{f}^k \otimes (\hat{b}_f^k (\hat{b}_f^k)^\dagger),
    \label{eq:v_f_reduced_rank}
\end{align} 
where $\otimes$ is the Kronecker product.

\subsection{Conventional Learning for LSTD-based Prediction}
\label{subsec:conven_lstd}
In conventional learning, the goal is to optimize the LSTD-based predictor $V_f^{(K)}$ by optimizing the feature matrix $\hat{B}_f$ and the feature-wise predictors $\{{v}_{f}^k\}_{k=1}^K$ based on the available training data set $\mathcal{Z}_f^\text{tr}$. Substituting $V_f$ with $V_f^{(K)}$ defined in \eqref{eq:v_f_reduced_rank} into the na\"ive extension of conventional learning in \eqref{eq:general_ridge_vec} yields the problem
\begin{align}
&V^{(K),*}(\mathcal{Z}_f^\text{tr} | \bar{V}^{(K)})= \argmin_{\substack{\hat{B}_f, v_f^{1},\ldots,v_f^{K}\\ V_f^{(K)} = \sum\limits_{k=1}^K v_{f}^k \otimes (\hat{b}_f^k (\hat{b}_f^k)^\dagger)}} \hspace{-0.5cm} || X_f^\text{tr} V_f^{(K)} - Y_f^\text{tr} ||_F^2 + \lambda \norm{V_f - \bar{V}^{(K)}}_F^2, \label{eq:general_ridge_vec_rr_naive_vbar}\nonumber\\[-10pt]
&\quad\quad\quad\quad\quad\quad\quad\quad\quad\quad\quad\quad\quad\quad\quad\quad\quad\quad\quad\quad\quad\quad\quad\quad\text{subject to } \hat{B}_f^\dagger \hat{B}_f = I_K,
\end{align}
over the optimization variables $(\hat{B}_f, \{ v_f^k \}_{k=1}^K)$. In \eqref{eq:general_ridge_vec_rr_naive_vbar}, the hyperparameters $(\lambda, \bar{V}^{(K)})$ are given by the scalar $\lambda > 0$ and by the $SN \times S$ LSTD-based bias matrix $\bar{V}^{(K)}$ defined as (cf. \eqref{eq:v_f_reduced_rank})
\begin{align}
    \bar{V}^{(K)} = \sum_{k=1}^K {\bar{v}}^k \otimes (\bar{b}^k (\bar{b}^k)^\dagger).
    \label{eq:v_f_reduced_rank_bias}
\end{align} 
Since the Euclidean norm regularization $\norm{V_f - \bar{V}^{(K)}}_F^2$ in \eqref{eq:general_ridge_vec_rr_naive_vbar} mixes long-term and short-term dependencies due to \eqref{eq:v_f_reduced_rank} and \eqref{eq:v_f_reduced_rank_bias}, we propose the modification of problem \eqref{eq:general_ridge_vec_rr_naive_vbar}
\begin{align}
&V^{(K),*}(\mathcal{Z}_f^\text{tr} | 
\{ \bar{b}^k, \bar{v}^k\}_{k=1}^K) \nonumber\\&= \hspace{-0.3cm}\argmin_{\substack{\hat{B}_f, v_f^{1},\ldots,v_f^{K}\\ V_f^{(K)} = \sum\limits_{k=1}^K v_{f}^k \otimes (\hat{b}_f^k (\hat{b}_f^k)^\dagger)}} \hspace{-0.5cm}\Big\{ || X_f^\text{tr} V_f^{(K)} - Y_f^\text{tr} ||_F^2  - \lambda_1 \sum_{k=1}^K \text{tr} \left((\hat{b}_f^k)^\dagger (\bar{b}^k (\bar{b}^k)^\dagger)  \hat{b}_f^k\right)  + \lambda_2 \sum_{k=1}^K|| v_f^k - \bar{v}^k ||^2  \Big\}, \label{eq:general_ridge_vec_rr}\nonumber\\
&\quad\quad\quad\quad\quad\quad\quad\quad\quad\quad\quad\quad\quad\quad\quad\quad\quad\quad\quad\quad\quad\quad\quad\quad\quad\quad\quad \text{subject to } \hat{B}_f^\dagger \hat{B}_f = I_K,
\end{align}
with hyperparameters $(\lambda_1, \lambda_2, \bar{b}^1, \ldots, \bar{b}^K, \bar{v}^{1},\ldots, \bar{v}^{K})$ given by the scalars $\lambda_1, \lambda_2 >0$, by the $S\times1$ long-term bias vectors $\bar{b}^1, \ldots, \bar{b}^K$, and by the $N\times 1$ short-term bias vectors $\bar{v}^1,\ldots,\bar{v}^K$. For each feature $k$, the considered regularization minimizes the Euclidean distance between the short-term prediction vector $v_f^k$ and the short-term bias vector $\bar{v}^k$ as in Sec.~\ref{sec:single_angenna}, while  maximizing the alignment between the long-term feature vector $\hat{b}_f^k$ and the long-term bias vector $\bar{b}^k$ in a manner akin to the kernel alignment method of \cite{cortes2012algorithms}. 


To address problem \eqref{eq:general_ridge_vec_rr}, inspired by \cite{wold1987principal, wong1999unified}, we propose a sequential approach, in which the pair $(v_f^k, \hat{b}_f^k)$ consisting of the $k$-th predictor $v_f^k$ and the $k$-th feature vector $\hat{b}_f^k$ is optimized in the order $k=1,2,\ldots,K$.
Specifically, at each step $k$, we consider the problem 
\begin{align}
\nonumber
\hat{b}_f^{k,*}, v_f^{k,*}&= \hspace{-0.7cm}\argmin_{\substack{\hat{b}_f^k, {v}_f^{k}\\ (V_f^{(K)})^k = {v}_{f}^k \otimes (\hat{b}_f^k (\hat{b}_f^k)^\dagger)}} \hspace{-0.6cm}\Big\{ || X_f^\text{tr} (V_f^{(K)})^k - (Y_f^\text{tr})^k ||_F^2 - \lambda_1 \text{tr} \left( (\hat{b}_f^k)^\dagger (\bar{b}^k (\bar{b}^k)^\dagger) \hat{b}_f^k \right)+ \lambda_2 || {v}_f^k - \bar{v}^k ||^2 \Big\}, \nonumber\\
&\quad\quad\quad\quad\quad\quad\quad\quad\quad\quad\quad\quad\quad\quad\quad\quad\quad\quad\quad\quad\quad\quad\quad\quad\quad\text{subject to } (\hat{b}_f^k)^\dagger \hat{b}_f^k = 1,\label{eq:general_ridge_vec_rr_seq}
\end{align}
where the $L^\text{tr} \times S$ \emph{$k$-th residual} target matrix  $(Y_f^\text{tr})^k$ is defined as \cite{wold1987principal, wong1999unified}
\begin{align}
    (Y_f^\text{tr})^k = \begin{cases}
Y_f^\text{tr}, \text{ for } k=1, \\
Y_f^\text{tr} - \sum\limits_{k'=1}^{k-1} X_f^\text{tr} (V_f^{(K)})^{k',*}, \text{ for } k > 1,
\end{cases}
\label{eq:conven_resi_target}
\end{align}
given the \emph{k-th predictor}
\begin{align}
(V_f^{(K)})^{k} = {v}_{f}^{k} \otimes (\hat{b}_f^{k} (\hat{b}_f^{k})^\dagger)
\end{align}
and \emph{k-th optimized predictor}
\begin{align}
(V_f^{(K)})^{k,*} = {v}_{f}^{k,*} \otimes (\hat{b}_f^{k,*} (\hat{b}_f^{k,*})^\dagger).
\end{align}

Since \eqref{eq:general_ridge_vec_rr_seq} is a nonconvex problem, we consider alternating least squares (ALS) \cite{sidiropoulos2017tensor} to obtain the optimal solution $\{\hat{b}_f^{k,*}, v_f^{k,*}\}$ by iterating between the following steps: \emph{(i)} for a fixed $\hat{b}_f^k$, update $v_f^k$ as
\begin{align}
    v_f^{k} \leftarrow \hspace{-0.85cm} \argmin_{\substack{{v}_f^{k}\\ (V_f^{(K)})^k = {v}_{f}^k \otimes (\hat{b}_f^k (\hat{b}_f^k))^\dagger}} \hspace{-0.85cm}\Big\{ || X_f^\text{tr} (V_f^{(K)})^k - (Y_f^\text{tr})^k ||_F^2 + \lambda_2 || {v}_f^k - \bar{v}^k ||^2 \Big\}; \label{eq:als_w}
\end{align}
and \emph{(ii)} for a fixed $v_f^k$, update $\hat{b}_f^k$ as
\begin{align}
    \hat{b}_f^{k} &\leftarrow \hspace{-0.7cm}\argmin_{\substack{\hat{b}_f^k\\ (V_f^{(K)})^k = {v}_{f}^k \otimes (\hat{b}_f^k (\hat{b}_f^k)^\dagger)}} \hspace{-0.75cm}\Big\{ || X_f^\text{tr} (V_f^{(K)})^k - (Y_f^\text{tr})^k ||_F^2 - \lambda_1 \text{tr} \left( (\hat{b}_f^k)^\dagger (\bar{b}^k (\bar{b}^k)^\dagger) \hat{b}_f^k \right) \Big\}, \label{eq:als_b}\nonumber\\
&\quad\quad\quad\quad\quad\quad\quad\quad\quad\quad\quad\quad\quad\quad\quad\quad\quad\quad\quad\quad\text{subject to } (\hat{b}_f^k)^\dagger \hat{b}_f^k = 1,
\end{align}
until convergence. Closed-form solutions for \eqref{eq:als_w} and \eqref{eq:als_b} can be found in Appendix~\ref{app:lstd_conven}, and the overall LSTD-based conventional learning scheme can be found in Algorithm.~\ref{alg:vec_conven_learning}.

\begin{algorithm}[h] 
\DontPrintSemicolon
\smallskip
\KwIn{training data set $\mathcal{Z}_f^\text{tr}$ at current frame $f$; number of total features $K$, hyperparameter $\lambda_1 > 0$, $\lambda_2 > 0$, $\bar{b}^1, \ldots, \bar{b}^K$, $\bar{v}^1, \ldots, \bar{v}^K$}
\KwOut{learned LSTD-based predictor $V_f^{(K),*}$}
\vspace{0.2cm}
\hrule
\vspace{0.2cm}
\texttt{LSTD-based conventional learning}: \\
initialize residual target matrix $(Y_f^\text{tr})^{k=1} \leftarrow Y_f^\text{tr} $ \\
\For{\em each feature $k=1\ldots,K$}{
$v_f^{k,*}, \hat{b}_f^{k,*}$ = \texttt{ALS}$(k, X_f^\text{tr}, (Y_f^\text{tr})^k, \lambda_1, \lambda_2, \bar{b}^k, \bar{v}^k)$ \\
update residual target matrix $(Y_f^\text{tr})^{k+1} \leftarrow  (Y_f^\text{tr})^{k} - X_f^\text{tr} v_f^k \otimes (\hat{b}_f^k (\hat{b}_f^k)^\dagger)$
}
\textbf{return} $V_f^{(K),*} = \sum_{k=1}^K v_f^k \otimes (\hat{b}_f^k (\hat{b}_f^k)^\dagger)$

\texttt{ALS}$(k, X_f^\text{tr}, (Y_f^\text{tr})^k, \lambda_1, \lambda_2, \bar{b}^k, \bar{v}^k)$: \\
\While{\em not converged}{
first update $v_f^k$ via \eqref{eq:als_w} then, update $\hat{b}_f^k$ via \eqref{eq:als_b}
}
return $v_f^k, \hat{b}_f^k$

\caption{LSTD-based conventional learning for channel prediction for $S \geq 1$}
\label{alg:vec_conven_learning}
\end{algorithm}

\subsection{Transfer Learning for LSTD-based Prediction}
Similar to conventional learning, transfer learning for LSTD-based prediction can be addressed from the na\"ive extension \eqref{eq:transfer_obj_vec_naive} by utilizing the LSTD parametrization $V^{(K)}$ in \eqref{eq:v_f_reduced_rank} in lieu of the unconstrained predictor $V$ to obtain the bias matrix $\bar{V}^{(K),\text{trans}}$ as
\begin{align}
    \bar{V}^{(K),\text{trans}} = \hspace{-0.6cm}\argmin_{\substack{\hat{B}, v^{1},\ldots,v^{K}\\ V^{(K)} = \sum\limits_{k=1}^K v^k \otimes (\hat{b}^k (\hat{b}^k)^\dagger)}} \hspace{-0.6cm}\Big\{ \sum_{f=1}^F || X_f V^{(K)} - Y_f ||_F^2  \Big\}, \label{eq:transfer_obj_vec_LSTD}\nonumber\\
\text{subject to } \hat{B}^\dagger \hat{B} = I_K,
\end{align}
which can also be solved via the ALS-based sequential approach detailed in Sec.~\ref{subsec:conven_lstd}. This produces the sequences $\bar{b}^{1,\text{trans}},\ldots,\bar{b}^{K,\text{trans}}$ and $\bar{v}^{1,\text{trans}},\ldots,\bar{v}^{K,\text{trans}}$ as (cf. \eqref{eq:general_ridge_vec_rr_seq})
\begin{align}
\nonumber
\bar{b}^{k,\text{trans}}, \bar{v}^{k,\text{trans}}= \hspace{-0.75cm}\argmin_{\substack{\hat{b}^k, {v}^{k}\\ (V^{(K)})^k = {v}^k \otimes (\hat{b}^k (\hat{b}^k)^\dagger)}} \hspace{-0.52cm}&\Big\{ \sum_{f=1}^F || X_f (V^{(K)})^k - (Y_f)^k ||_F^2  \Big\}, \\
&\quad\text{subject to } (\hat{b}^k)^\dagger \hat{b}^k = 1,\label{eq:general_ridge_vec_rr_seq_trans}
\end{align}
where the residual target matrix $(Y_f)^k$ is defined as (cf. \eqref{eq:conven_resi_target})
\begin{align}
(Y_f)^k = \begin{cases}
Y_f, \text{ for } k=1, \\
Y_f - \sum\limits_{k'=1}^{k-1} X_f (V_f^{(K)})^{k',\text{trans}}, \text{ for } k > 1
\end{cases} 
\label{eq:meta_resi_target_te}
\end{align}
with $k$-th optimized predictor
\begin{align}
    (V^{(K)})^{k,\text{trans}} = \bar{v}^{k,\text{trans}} \otimes (\bar{b}^{k,\text{trans}} (\bar{b}^{k,\text{trans}}))^\dagger.
\end{align}
Details for transfer learning can be found in Appendix~\ref{app:lstd_transfer}, and the overall transfer learning scheme for LSTD-prediction is summarized in  Algorithm~\ref{alg:vec_transfer_learning}. After transfer learning, similar to Sec.~\ref{sec:scalar_transfer}, based on the optimized hyperparameters $\bar{b}^{1,\text{trans}},\ldots,\bar{b}^{K,\text{trans}}$ and $\bar{v}^{1,\text{trans}},\ldots,\bar{v}^{K,\text{trans}}$, LSTD-based channel predictor for a new frame $f^\text{new}$ can be obtained via \eqref{eq:general_ridge_vec_rr} as 
\begin{align}
V^{(K),*}_{f^\text{new}} = V^{(K),*}\big(\mathcal{Z}_{f^\text{new}}^\text{tr} | \{ \bar{b}^{k,\text{trans}}, \bar{v}^{k,\text{trans}}\}_{k=1}^K\big),
\end{align}
which can also be solved in the sequential way as in \eqref{eq:general_ridge_vec_rr_seq}.

\begin{algorithm}[h] 
\DontPrintSemicolon
\smallskip
\KwIn{previous data  $\{\{ h_{l,f} \}_{l=1}^{L+N+\delta-1}\}_{f=1}^F$; window size $N$; prediction lag size $\delta$; number of total features $K$}
\KwOut{transfer-learned hyperparameters $\bar{b}^{1,\text{trans}},\ldots,\bar{b}^{K,\text{trans}}, \bar{v}^{1,\text{trans}}, \ldots,\bar{v}^{K,\text{trans}} $}
\vspace{0.2cm}
\hrule
\vspace{0.2cm}
initialize $X^\text{trans}$ and $Y^\text{trans}$ to empty matrices \\
\For{{\em each frame $f \in \{1,\ldots,F\}$}}{
$X^\text{trans} = [X^\text{trans} , X_f]^\top$ and $Y^\text{trans} = [Y^\text{trans} , Y_f]^\top$
}
follow \emph{LSTD-based conventional learning} in Algorithm~\ref{alg:vec_conven_learning} with input  $\mathcal{Z}_f^\text{tr} = (X^\text{trans}, Y^\text{trans}), \lambda_1 = 0, \lambda_2 = 0$

\caption{LSTD-based transfer-learning for channel prediction for $S \geq 1$}
\label{alg:vec_transfer_learning}
\end{algorithm}

\subsection{Meta-Learning for LSTD-based Prediction}
\label{sec:meta_lstd}
Plugging \eqref{eq:general_ridge_vec_rr} to na\"ive extension of \eqref{eq:meta_obj_vec_naive}, we can formulate meta-learning problem for LSTD-based prediction as 
\begin{align}
\min_{\{\bar{b}^k, \bar{v}^k\}_{k=1}^K}  \hspace{-0.1cm}\sum_{f=1}^F \norm{ X_f^\text{te} V^{(K),*}(\mathcal{Z}_f^\textrm{tr}|\{ \bar{b}^{k}, \bar{v}^{k}\}_{k=1}^K) - Y_{f}^\textrm{te}}_F^2.
\label{eq:meta_obj_vec_lstd}
\end{align}
Similar to the sequential approach \eqref{eq:general_ridge_vec_rr_seq} described in Sec.~\ref{subsec:conven_lstd}, we propose a \emph{hierarchical sequential} approach for meta-learning using \eqref{eq:general_ridge_vec_rr_seq} in the order $k=1,\ldots,K$, obtaining the problem
\begin{align}
\bar{b}^{k,\text{meta}}, \bar{v}^{k,\text{meta}}= \hspace{-1.3cm}\argmin_{\substack{\bar{b}^k, \bar{v}^k \\ (V_f^{(K)})^{k,*} = v_f^{k,*} \otimes (\hat{b}_f^{k,*} (\hat{b}_f^{k,*})^\dagger) }} \hspace{-1.3cm}\Big\{ \sum_{f=1}^F \norm{  X_f^\text{te} (V_f^{(K)})^{k,*} - (Y_{f}^\text{te})^k}_F^2  \Big\}, \label{eq:general_ridge_vec_rr_seq_meta}
\end{align}
with the residual target matrix $(Y_f^\text{te})^k$ defined as (cf. \eqref{eq:conven_resi_target})
\begin{align}
(Y_f^\text{te})^k = \begin{cases}
Y_f^\text{te}, \text{ for } k=1, \\
Y_f^\text{te} - \sum\limits_{k'=1}^{k-1} X_f^\text{te} (V_f^{(K)})^{k',*}, \text{ for } k > 1.
\end{cases} 
\label{eq:meta_resi_target_te}
\end{align}


The bilevel non-convex optimization problem \eqref{eq:general_ridge_vec_rr_seq_meta} is addressed through gradient-based updates with gradients computed via equilibrium propagation (EP) \cite{scellier2017equilibrium, zucchet2021contrastive}. EP uses finite differentiation to approximate the gradient of the bilevel optimization \eqref{eq:general_ridge_vec_rr_seq_meta}, where the difference is computed between two gradients obtained at two stationary points $(\hat{b}_f^{k,*}, v_f^{k,*})$ and $(\hat{b}_f^{k,\alpha}, v_f^{k,\alpha})$ for the original problem \eqref{eq:general_ridge_vec_rr_seq} and modified version of \eqref{eq:general_ridge_vec_rr_seq} that considers additional prediction loss for the test set $\mathcal{Z}_f^\text{te}$. Specifically, EP leverages the asymptotic equality \cite{scellier2017equilibrium} 
\begin{align}
    \nabla_{\bar{b}^k}&\sum_{f=1}^F \norm{  X_f^\text{te} (V_f^{(K)})^{k,*} - (Y_{f}^\text{te})^k}_F^2 =  \lim_{\alpha \rightarrow 0} \frac{2\lambda_1}{\alpha} \sum_{f=1}^F \left( \hat{b}_f^{k,*} (\hat{b}_{f}^{k,*})^\dagger - \hat{b}_{f}^{k,\alpha} (\hat{b}_{f}^{k,\alpha})^\dagger \right)\bar{b}^k
    \label{eq:ep_grad_b}
\end{align}
and
\begin{align}
    \nabla_{\bar{v}^k}\sum_{f=1}^F &\norm{  X_f^\text{te} (V_f^{(K)})^{k,*} - (Y_{f}^\text{te})^k}_F^2  =  \lim_{\alpha \rightarrow 0} \frac{2\lambda_2}{\alpha} \sum_{f=1}^F(v^{k,*}_f - v^{k,\alpha}_f),
    \label{eq:ep_grad_v}
\end{align}
with additional real-valued hyperparameter $\alpha \in \mathbb{R}$ which is generally chosen to be a non-zero small value \cite{scellier2017equilibrium, zucchet2021contrastive}. In \eqref{eq:ep_grad_v}--\eqref{eq:ep_grad_b}, vectors $\hat{b}_f^{k,\alpha}$ and $v_f^{k,\alpha}$ are defined as (cf. \eqref{eq:general_ridge_vec_rr_seq})
\begin{align}
\nonumber
\hat{b}_f^{k,\alpha}, v_f^{k,\alpha} &= \hspace{-0.8cm}\argmin_{\substack{\hat{b}_f^k, {v}_f^{k}\\ (V_f^{(K)})^k = {v}_{f}^k \otimes (\hat{b}_f^k (\hat{b}_f^k)^\dagger)}} \hspace{-0.8cm}\Big\{ || X_f^\text{tr} (V_f^{(K)})^k - (Y_f^\text{tr})^k ||_F^2 + \alpha || X_f^\text{te} (V_f^{(K)})^k - (Y_f^\text{te})^k ||_F^2  \nonumber\\&\quad\quad\quad\quad\quad\quad\quad\quad\quad\quad- \lambda_1 \text{tr} \left( (\hat{b}_f^k)^\dagger (\bar{b}^k (\bar{b}^k)^\dagger) \hat{b}_f^k \right)+ \lambda_2 || {v}_f^k - \bar{v}^k ||^2 \Big\}, \nonumber\\
&\quad\quad\quad\quad\quad\quad\quad\quad\quad\quad\quad\quad\quad\quad\quad\quad\quad\quad\quad\text{subject to } (\hat{b}_f^k)^\dagger \hat{b}_f^k = 1.\label{eq:general_ridge_vec_rr_seq_meta_alpha}
\end{align}
Derivations for the gradients \eqref{eq:ep_grad_b} and \eqref{eq:ep_grad_v} can be found in Appendix~\ref{app:lstd_meta}.

To reduce the computational complexity for the gradient-based updates, we adopt stochastic gradient descent with the Adam optimizer as done in \cite{zucchet2021contrastive} in order to update $\bar{b}^k$ and $\bar{v}^k$ based on \eqref{eq:ep_grad_b}--\eqref{eq:ep_grad_v}. The overall LSTD-based meta-learning scheme is detailed in Algorithm~\ref{alg:vec_meta_learning}.

After meta-learning, as in Sec.~\ref{sec:scalar_meta}, based on the optimized $\bar{b}^{1,\text{meta}},\ldots,\bar{b}^{K,\text{meta}}$ and $\bar{v}^{1,\text{meta}},\ldots,\bar{v}^{K,\text{meta}}$, LSTD-based channel predictor for a new frame $f^\text{new}$ can be obtained via \eqref{eq:general_ridge_vec_rr} as 
\begin{align}
V^{(K),*}_{f^\text{new}} = V^{(K),*}(\mathcal{Z}_{f^\text{new}}^\text{tr} | \{ \bar{b}^{k,\text{meta}}, \bar{v}^{k,\text{meta}}\}_{k=1}^K)
\label{eq:meta_test_v}
\end{align}
which can be solved in sequential way as in \eqref{eq:general_ridge_vec_rr_seq}.
\begin{algorithm}[h] 
\DontPrintSemicolon
\smallskip
\KwIn{previous data  $\{\{ h_{l,f} \}_{l=1}^{L+N+\delta-1}\}_{f=1}^F$; training set size $L^\textrm{tr}$; test set size $L^\textrm{te} = L-L^\textrm{tr}$; window size $N$; prediction lag size $\delta$; number of total features $K$; step size for gradient-based update $\kappa$; hyperparameter $\lambda_1 > 0, \lambda_2 > 0, \alpha \ll 1$}
\KwOut{meta-learned hyperparameters $\bar{b}^{1,\text{meta}},\ldots,\bar{b}^{K,\text{meta}}, \bar{v}^{1,\text{meta}}, \ldots,\bar{v}^{K,\text{meta}} $}
\vspace{0.2cm}
\hrule
\vspace{0.2cm}
initialize hyperparameters $\bar{b}^1, \ldots, \bar{b}^K, \bar{v}^1,\ldots,\bar{v}^K$ \\
\For{{\em each frame $f \in \{1,\ldots,F\}$}}{
split $\mathcal{Z}_f$ into training set $\mathcal{Z}_f^\text{tr}$ and test set $\mathcal{Z}_f^\text{te}$ \\
initialize residual target matrices $(Y_f^\text{tr})^{k=1} \leftarrow Y_f^\text{tr} $, $(Y_f^\text{te})^{k=1} \leftarrow Y_f^\text{te} $ \\
}
\For{\em each feature $k=1,\ldots,K$}{
\hspace{-0.2cm}\texttt{EP-based meta-learning for feature $k$}\\
\While{\em not converged}{
\For{\em each frame $f \in \{1,\ldots,F \}$}{
$v_f^{k,*}, \hat{b}_f^{k,*}$ = \texttt{ALS}$(k, X_f^\text{tr}, (Y_f^\text{tr})^k, \lambda_1, \lambda_2, \bar{b}^k, \bar{v}^k)$  \\
$v_f^{k,\alpha}, \hat{b}_f^{k,\alpha}$ = \texttt{ALS}$(k, [X_f^\text{tr}, \sqrt{\alpha} X_f^\text{te}]^\top,[(Y_f^\text{tr})^k, \sqrt{\alpha} (Y_f^\text{te})^k]^\top,\lambda_1, \lambda_2, \bar{b}^k, \bar{v}^k)$}
$\bar{v}^k \leftarrow  \bar{v}^k - \kappa \frac{2\lambda_2}{\alpha} \sum_{f=1}^F(v^{k,*}_f - v^{k,\alpha}_f) $ \\
$\bar{b}^k \leftarrow \bar{b}^k-\kappa  \frac{2\lambda_1}{\alpha} \sum_{f=1}^F \left( b_f^{k,*} (b_{f}^{k,*})^\dagger - b_{f}^{k,\alpha} (b_{f}^{k,\alpha})^\dagger \right)\bar{b}^k$
}
$\bar{v}^{k,\text{meta}} \leftarrow \bar{v}^k$, $\bar{b}^{k,\text{meta}} \leftarrow \bar{b}^k$ \\
update residual target matrix for both training and test set $(Y_f)^{k+1} \leftarrow  (Y_f)^{k} - X_f v_f^{k,*} \otimes (\hat{b}_f^{k,*} (\hat{b}_f^{k,*})^\dagger)$  for every frame $f$
}

\caption{LSTD-based meta-learning for channel prediction for $S \geq 1$}
\label{alg:vec_meta_learning}
\end{algorithm}

\subsection{Rank-Estimation for LSTD-based Prediction}
\label{sec:rank_estimation}



Number of total features $K$ for LSTD-based predictions depends on the rank of the unknown space-time signature matrix $T_f$ as discussed in Sec.~\ref{sec:subsec_LSTD}. This rank can be estimated by using available channels from previous frames if we assume that the number of total features does not change over multiple frames. This can be achieved via one of the standard methods, Akaike’s information theoretic criterion (AIC) (eq. (16) in \cite{wax1985detection}), which is applicable for all the proposed LSTD-based techniques. However, as AIC-based rank estimation is generally tend to be overestimated \cite{wax1985detection, liavas1999blind}, we  propose a potentially more effective estimator for meta-learning which utilizes validation data set.
 To this end, we  first split the available $F$ frames into $F^\text{tr}$ meta-training frames $f=1,\ldots,F^\text{tr}$ and $F^\text{val}$ meta-validation frames $f=F^\text{tr}+1,\ldots,F$. Then, we compute the sum-loss as (cf. \eqref{eq:meta_obj_vec_lstd})
\begin{align}
 \sum_{f=F^\text{tr}+1}^{F^\text{tr}+F^\text{val}} \Big\| X_f^\text{te} V^{(k),*}(\mathcal{Z}_f^\textrm{tr}|\{ \bar{b}^{k',\text{meta}}, \bar{v}^{k',\text{meta}}\}_{k'=1}^k) - Y_{f}^\textrm{te}\Big\|_F^2,
\label{eq:meta_obj_vec_lstd_val}
\end{align}
where the hyperparameters $\{ \bar{b}^{k',\text{meta}}, \bar{v}^{k',\text{meta}}\}_{k'=1}^k$ are computed using the $F^\text{tr}$ meta-training frames, as explained in the previous section.  The rank-estimation procedure sequentially evaluates the meta-validation loss \eqref{eq:meta_obj_vec_lstd_val} in order to minimize it over the selection of $k$. 
In this regards, it is worth noting that an increase in total number of features $k$ always decreases the meta-training loss in \eqref{eq:meta_obj_vec_lstd}, while this is not necessarily true for the meta-validation and meta-test losses.

\section{Experiments}
\label{sec:experiments}
In this section, we present experimental results for the prediction of multi-antenna and/or frequency-selective channels\footnote{\textcolor{black}{Code is} available at \url{https://github.com/kclip/channel-prediction-meta-learning}.}. Numerical examples for single-antenna frequency-flat channels for both offline and online learning scenarios can be found in the conference version of this paper \cite{park2021predicting}. For all the experiments, we compute the normalized mean squared error (NMSE) $||\hat{h}_{l+\delta,f} - {h}_{l+\delta,f}||^2 / ||h_{l+\delta,f}||^2 $, which is averaged over $100$ samples for $200$ new frames. To avoid discrepancies between the evaluation measures used during training and testing phase, we also adopt the NMSE as the training loss function by normalizing the training data set for the new frame $f^\text{new}$ as (cf. \eqref{eq:training_dataset})
\begin{align}
\mathcal{Z}^\text{tr}_{f^\text{new}} &= \{ (x_{i,{f^\text{new}}}, y_{i,{f^\text{new}}})\}_{i=1}^{L^\text{tr}}\equiv \Big\{\Big(\frac{\text{vec}(H_{l,{f^\text{new}}}^N)}{\norm{h_{l+\delta,{f^\text{new}}}}},  \frac{h_{l+\delta,{f^\text{new}}}}{\norm{h_{l+\delta,{f^\text{new}}}}}\Big)\Big\}_{l=N}^{L^\text{tr}+N-1},
\label{eq:training_dataset_nmse}
\end{align}
and similarly redefine the data sets from previous frames $f=1,\ldots,F$ for transfer and meta-training as (cf. \eqref{eq:Z_f_def})
\begin{align}
    \mathcal{Z}_f  &= \{(x_{i,f}, y_{i,f})\}_{i=1}^{L} \equiv \Big\{\Big(\frac{\text{vec}(H_{l,f}^N)}{\norm{h_{l+\delta,f}}}, \frac{h_{l+\delta,f}}{\norm{h_{l+\delta,f}}}\Big)\Big\}_{l=N}^{L+N-1}
    \label{eq:Z_f_def}.
\end{align}

As summarized in Table~\ref{table:exp_settings}, we consider a window size $N=5$ with lag size $\delta=3$. All of the experimental results follow the 3GPP 5G standard SCM channel model \cite{3gpp_tr_901} with variations of the long-term features over frames following Clause 7.6.3.2 (Procedure B) \cite{3gpp_tr_901}, under the Umi-Street Canyon environment, as discussed in Sec.~\ref{subsec:channel_model}. The normalized Doppler frequency 
{$\rho=\gamma_{d,f}/\gamma_{\text{SRS}} \in [0,1]$ within each frame $f$,  defined as the ratio between the Doppler frequency $\gamma_{d,f}$ \eqref{eq:multivariate_channel_model} and the frequency of the pilot symbols $\gamma_{\text{SRS}}$, or sounding reference signal (SRS) \cite{3gpp_tr_901}},  is randomly selected in one of the two following ways: \emph{(i)} for \emph{slow-varying environments}, it is uniformly drawn in the interval $[0.005, 0.05]$; and \emph{(ii)} for \emph{fast-varying environments}, it is uniformly distributed in the interval $[0.1, 1 ]$. 
In the following, we study the impact of \emph{(i)} the number of antennas $N_RN_T$; \emph{(ii)} the number of channel taps $W$; \emph{(iii)} the number of training samples $L^\text{new}$; and \emph{(iv)} the number of previous frames $F$, for various prediction schemes: (a) conventional learning; (b) transfer learning; and (c) meta-learning, where each scheme is implemented using either the na\"ive or the LSTD parametrization.

\begin{center}
\begin{table}[h!]
\caption{Experimental Setting}
\label{table:exp_settings}
\centering
\begin{tabular}{||c | c||} 
 \hline
 window size $(N)$ & $5$ \\ 
 \hline
 lag size $(\delta)$ & $3$ \\
 \hline
 number of previous frames $(F)$ & $500$ \\
 \hline
 number of slots $(L)$ & $100$ \\
 \hline
 frequency of the pilot signals ($w_{\text{SRS}}/2\pi$) & $200$ \\
 \hline
 normalized Doppler frequency  for slow-varying environment & $\rho \sim \text{Unif}[0.005, 0.05]$ \\
 \hline
  normalized Doppler frequency  for fast-varying environment  & $\rho \sim \text{Unif}[0.1, 1]$  \\
 \hline
  {\color{black} SNR for channel estimation}  &  {\color{black}$20$ dB}  \\
 \hline
   {\color{black} number of pilots for channel estimation}  &  {\color{black}$100$}  \\
 \hline
\end{tabular}
\end{table}
\end{center}

\subsection{Multi-Antenna Frequency-Flat Channels}
\label{sec:exp_per_antenna}
\begin{figure}
    \centering
    \hspace{-0.7cm}
    \includegraphics[width=0.5\columnwidth]{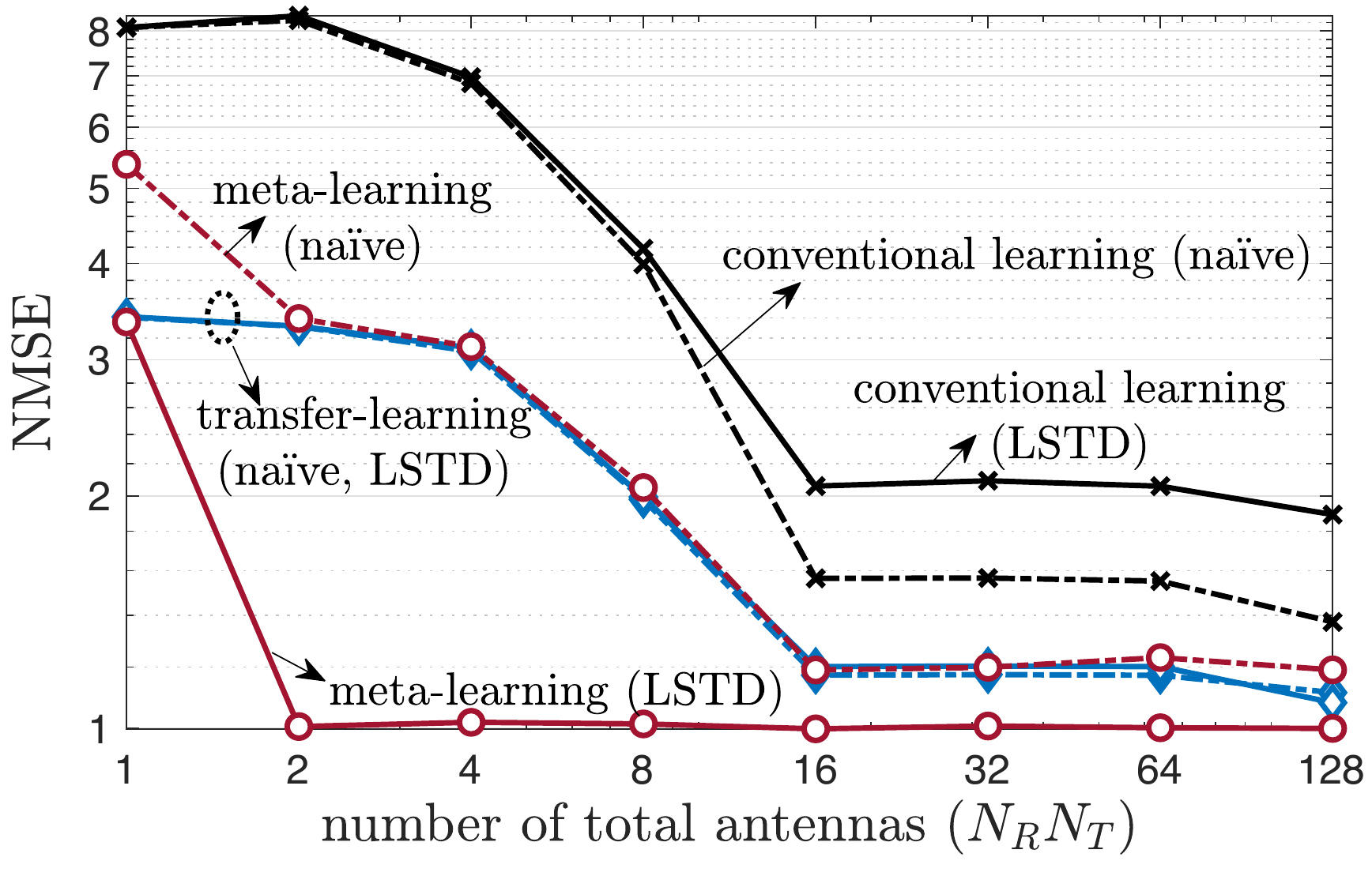}
    \caption{Multi-antenna frequency-flat channel prediction performance as a function of the total number of antennas, $N_RN_T$, under a single-clustered, single-tap ($W=1$), 3GPP SCM channel model for a fast-varying environment with  number of training samples $L^\text{new}=1$ ($K=1$).}
    \label{fig:vec_over_antenna_fast}
\end{figure}

\begin{figure}
    \centering
    \hspace{-0.7cm}
    \includegraphics[width=0.5\columnwidth]{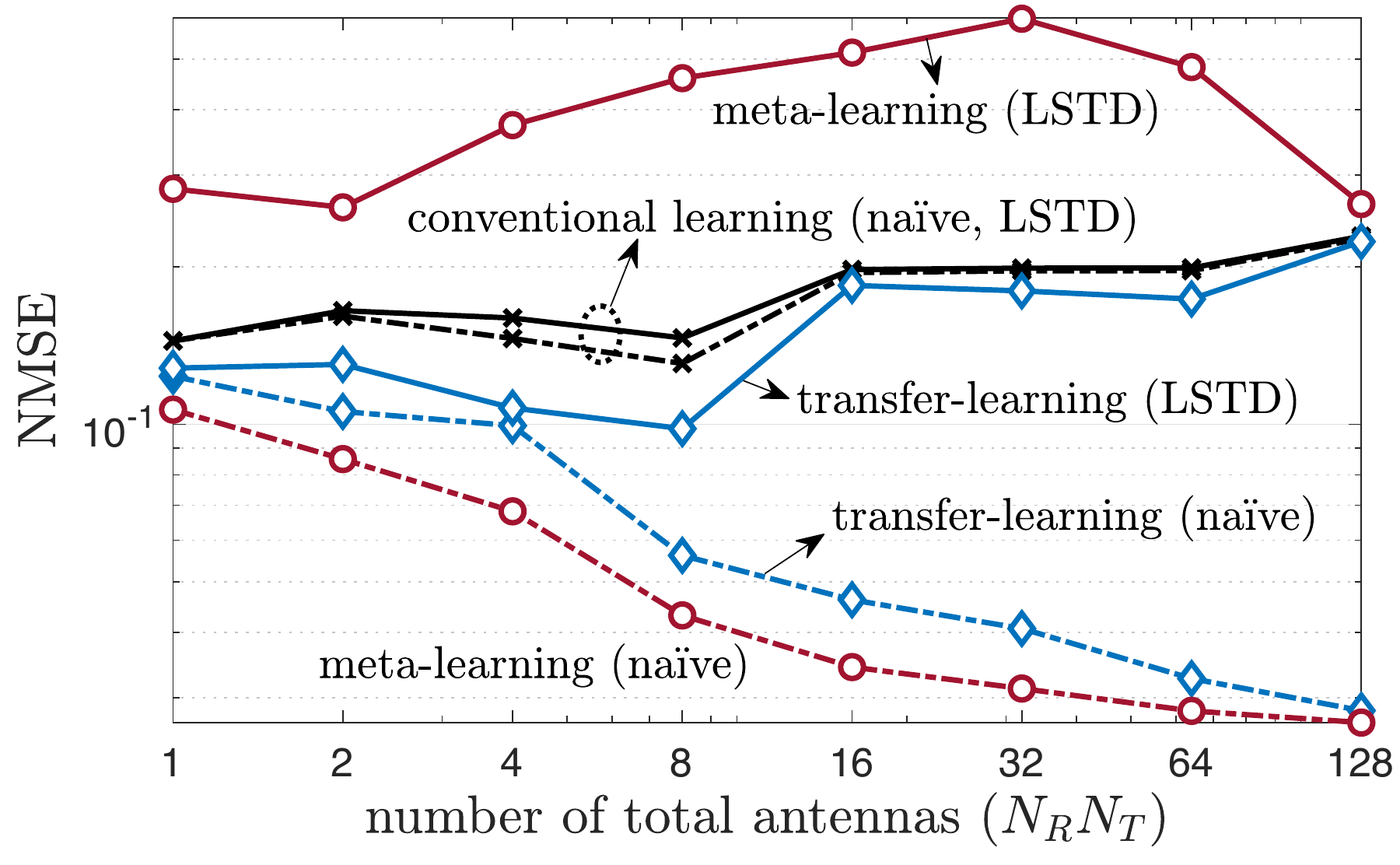}
    \caption{Multi-antenna frequency-flat channel prediction performance as a function of the total number of antennas, $N_RN_T$, under a single-clustered, single-tap ($W=1$), 3GPP SCM channel model for a slow-varying environment with  number of training samples $L^\text{new}=1$ ($K=1$).}
    \vspace{-0.5cm}
    \label{fig:vec_over_antenna_slow}
\end{figure}

We begin by considering multi-antenna frequency-flat channels and evaluating the NMSE as a function of total number of antennas $N_RN_T$ under a fast-varying environment (Fig.~\ref{fig:vec_over_antenna_fast}) or a slow-varying environment (Fig.~\ref{fig:vec_over_antenna_slow}). We set $K=1$ in the LSTD model. Specific antenna configuration are described in Appendix~\ref{app:per_antenna_exp}. 
Both transfer and meta-learning are seen to provide significant advantages as compared to conventional learning, as long as one chooses the type of parametrization -- na\"ive or LSTD -- as  a function of the type of variability in the channel, with meta-learning generally outperforming transfer learning. In particular, as seen in Fig. 4, for fast-varying environments, meta-learning with LSTD parametrization has the best performance, significantly reducing the NMSE with respect to both conventional and transfer learning. This is because meta-learning with LSTD can account for the need to adapt to fast-varying channel conditions, while also leveraging the reduced-rank structure of the channel.  In contrast, as shown in Fig. 5, for slow-varying channels, na\"ive parametrization tends to be preferable, since, as explained in Sec.  IV-B, long-term and short-term features of the channel become indistinguishable when channel variability is too low.  It is also interesting to observe that increasing the number of antennas is generally useful for prediction, as the predictor can build on a larger vector of correlated covariates. This is, however, not the case for conventional learning in slow-varying environments, for which the features tend to be too correlated, resulting in overfitting.

\subsection{Rank Estimation}
\label{sec:subsec_self_Rank}
\begin{figure}
    \centering
    \hspace{-0.7cm}
    \includegraphics[width=0.5\columnwidth]{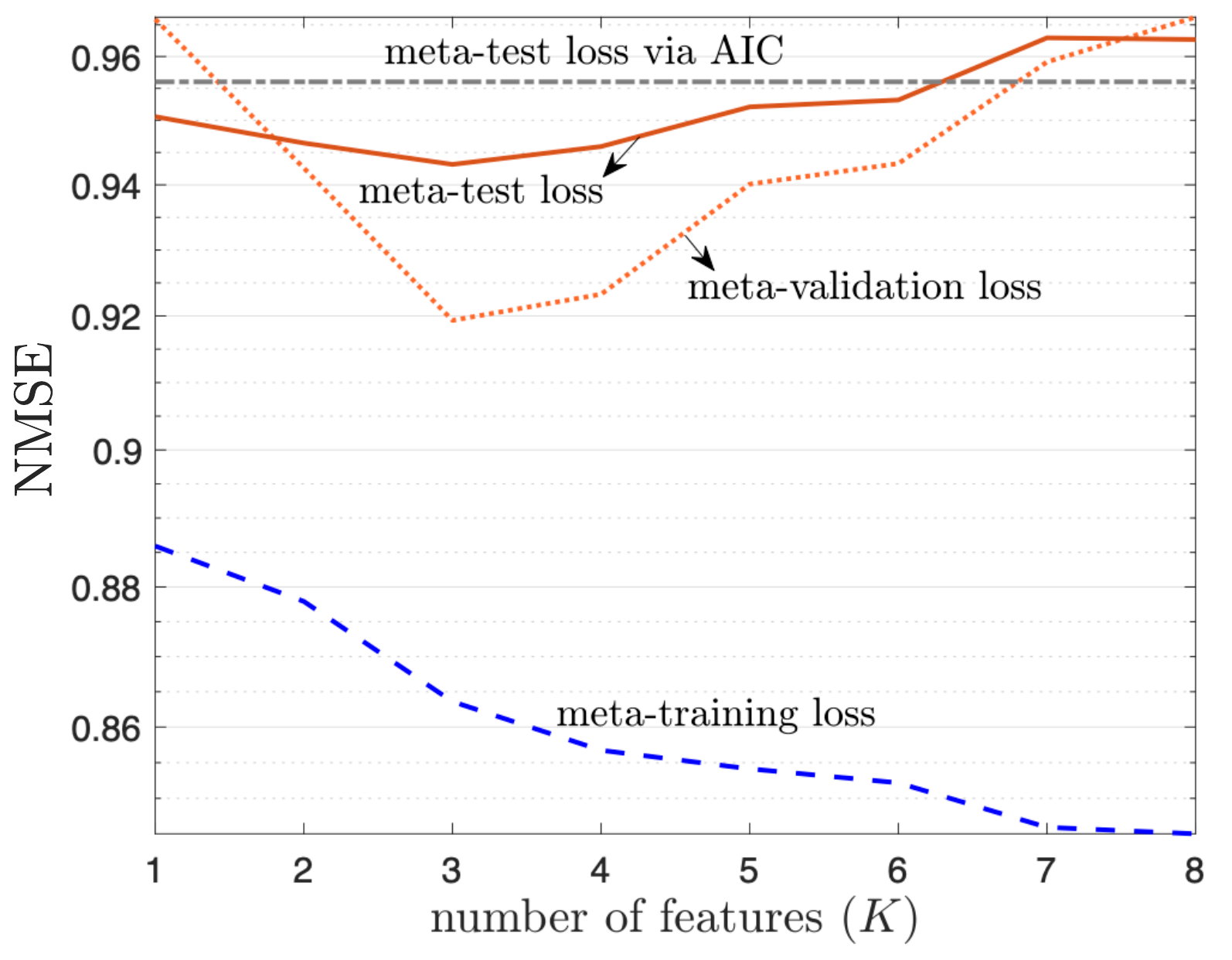}
    \caption{Multi-antenna frequency-selective channel prediction performance as a function of the number of features $K$, under $19$-clustered, multi-taps $(W=4)$, multi-antenna ($N_T=8, N_R=8$) 3GPP SCM channel model for a fast-varying environment with number of training samples $L^\text{new}=1$. Results are evaluated with number of previous frames $F^\text{tr}=20$ for meta-training; $F^\text{val}=20$ for meta-validation; and $F^\text{te}=200$ for meta-test.}
    \vspace{-0.5cm}
    \label{fig:vec_per_K}
\end{figure}

In the previous experiments, we have considered channels with unitary rank, for which one can assume without loss of optimality a number of features in the LSTD parametrization equal to $K=1$. In order to  implement predictors for multi-antenna
frequency-selective channels, instead, one needs to first address the problem of estimating the number of  features. Here, we evaluate the performance of the approach proposed in Sec.~\ref{sec:rank_estimation} for rank estimation. To this end, we set the number of antennas as $N_R = 8$ and $N_T = 8$, and consider the $19$-clustered channel model with delay spread ratio $2$. Fig.~\ref{fig:vec_per_K} shows the NMSE evaluated on the meta-training, meta-validation, and meta-test data sets as a function of total number of features $K$. The meta-training set contains $20$ frames, the meta-test $200$ frames, and the meta-validation set $20$ frames. The meta-training
loss is monotonically decreasing with $K$, since a richer parametrization enables a closer fit of the training data. In contrast, both meta-test and meta-validation loss are optimized for an intermediate value of $K$. The main point of the figure is that the meta-validation loss, while only containing $20$ frames, provides useful information to choose a value of $K$ that approximately minimizes the meta-test loss. In contrast, while we can see that $K=3$ is a proper estimate of the channel rank for the considered set-up, AIC-based rank estimation gives the highly overestimated value $K=200$, which deteriorates the prediction performance, as can be seen in Fig.~\ref{fig:vec_per_K}. 
Throughout the following experiments, we will follow the proposed procedure to select $K$ for meta-learning, while for all the other schemes, we adopt AIC-based rank estimation to determine $K$.

\subsection{Single-Antenna Frequency-Selective Channels}
\begin{figure}
    \centering
    \hspace{-0.7cm}
    \includegraphics[width=0.5\columnwidth]{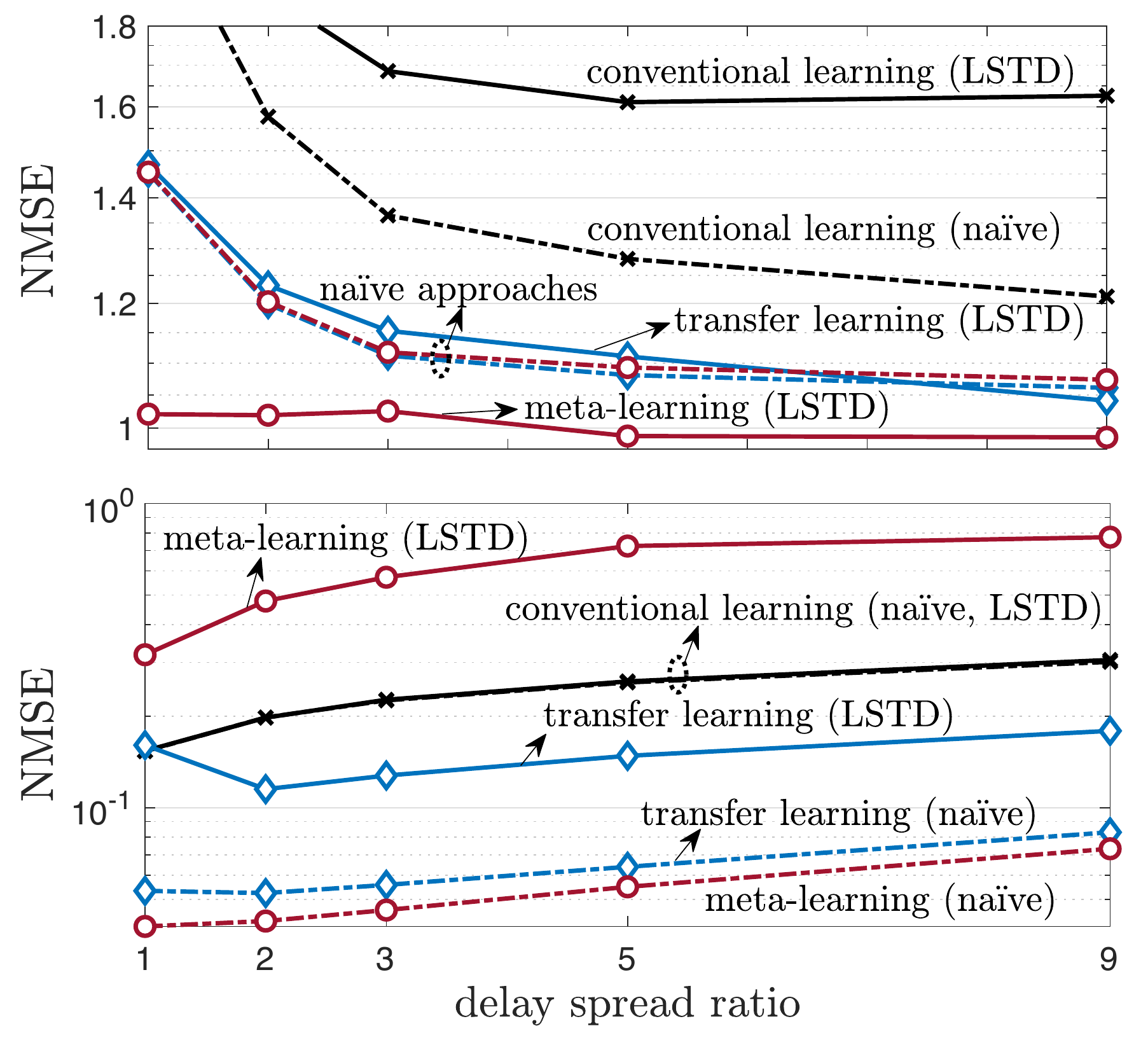}
    \caption{Single-antenna frequency-selective channel prediction performance as a function of delay spread ratio, under $19$-clustered, multi-taps, single-antenna ($N_T=1, N_R=1$) 3GPP SCM channel model for a fast-varying environment (top) and slow-varying environment (bottom) with number of training samples $L^\text{new}=1$ ($K=1$).}
    \label{fig:vec_per_W}
\end{figure}

{\color{black}Before considering multi-antenna frequency-selective channels, we first} consider the impact of the level of frequency selectivity on the prediction of single-antenna frequency-selective channels. To this end,  starting from $45\text{ ns}$, we increase the delay spread by a multiplicative factor,  and correspondingly also increase the number of taps by the same amount, which is referred to as delay spread ratio in Fig. 6. The number of taps $W$ is obtained as the smallest number of taps that contains more than 90\% of
the average channel power, following ITU-R report \cite{recommendationsmultipath}. Fig.~\ref{fig:vec_per_W} shows that the dependence on the delay spread of the channel is qualitatively similar to the dependence on the number of antennas in Fig.~\ref{fig:vec_over_antenna_fast} and Fig.~\ref{fig:vec_over_antenna_slow}, with the top of Fig.~\ref{fig:vec_per_W} representing the performance under a fast-varying environment and the bottom figure depicting the NMSE for a slow-varying environment. Accordingly, as discussed in the previous subsection, meta-learning outperforms both transfer and conventional learning, as long as the parametrization is correctly selected: na\"ive for slow-varying channels, and LSTD for fast-varying environments.

\subsection{Multi-Antenna Frequency-Selective Channel Case}
\begin{figure}
    \centering
    \hspace{-0.7cm}
    \includegraphics[width=0.5\columnwidth]{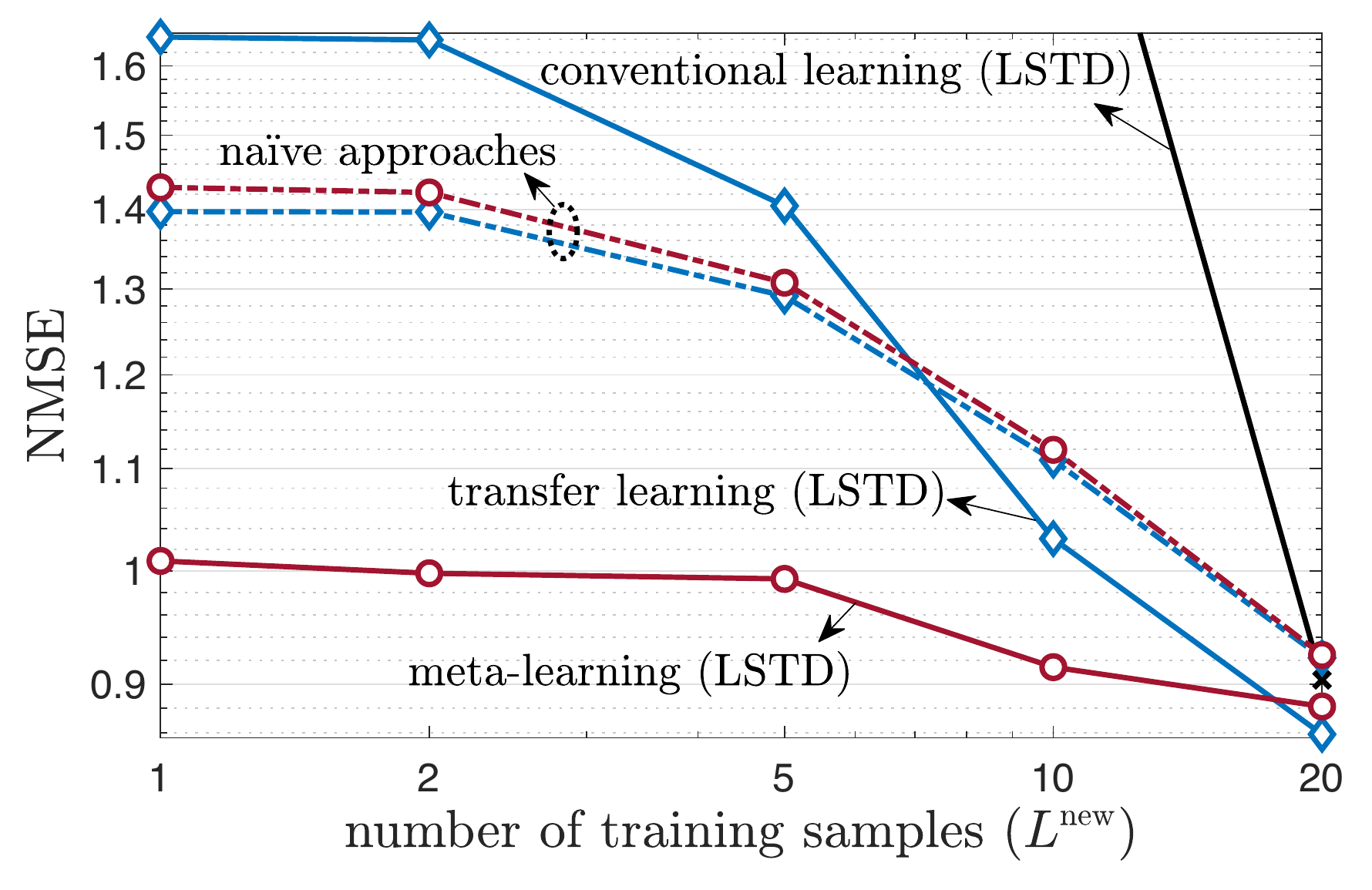}
    \caption{Multi-antenna frequency-selective channel prediction performance as a function of the number of training samples $L^\text{new}$, under $19$-clustered, two taps ($W=2$), multi-antenna ($N_T=4, N_R=2$) 3GPP SCM channel model for a fast-varying environment with total number of features $K=2$ unless determined by Sec.~\ref{sec:subsec_self_Rank}}.
    \label{fig:vec_per_L_fast}
    \vspace{-0.5cm}
\end{figure}

\begin{figure}
    \centering
    \hspace{-0.7cm}
    \includegraphics[width=0.5\columnwidth]{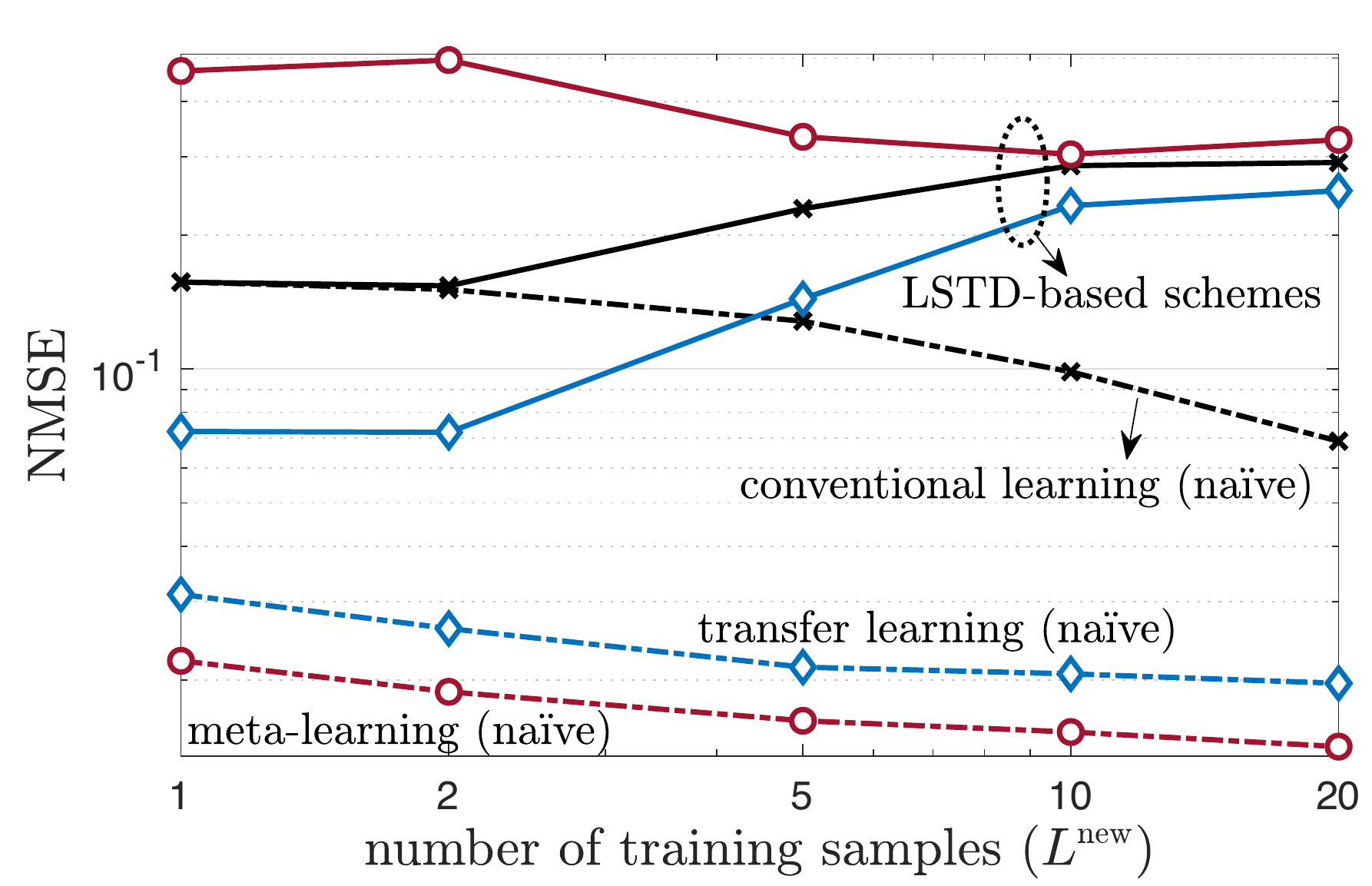}
    \caption{Multi-antenna frequency-selective channel prediction performance as a function of the number of training samples $L^\text{new}$, under $19$-clustered, two taps ($W=2$), multi-antenna ($N_T=4, N_R=2$) 3GPP SCM channel model for a slow-varying environment.}.
    \label{fig:vec_per_L_slow}
\end{figure}

\begin{figure}
    \centering
    \hspace{-0.7cm}
    \includegraphics[width=0.5\columnwidth]{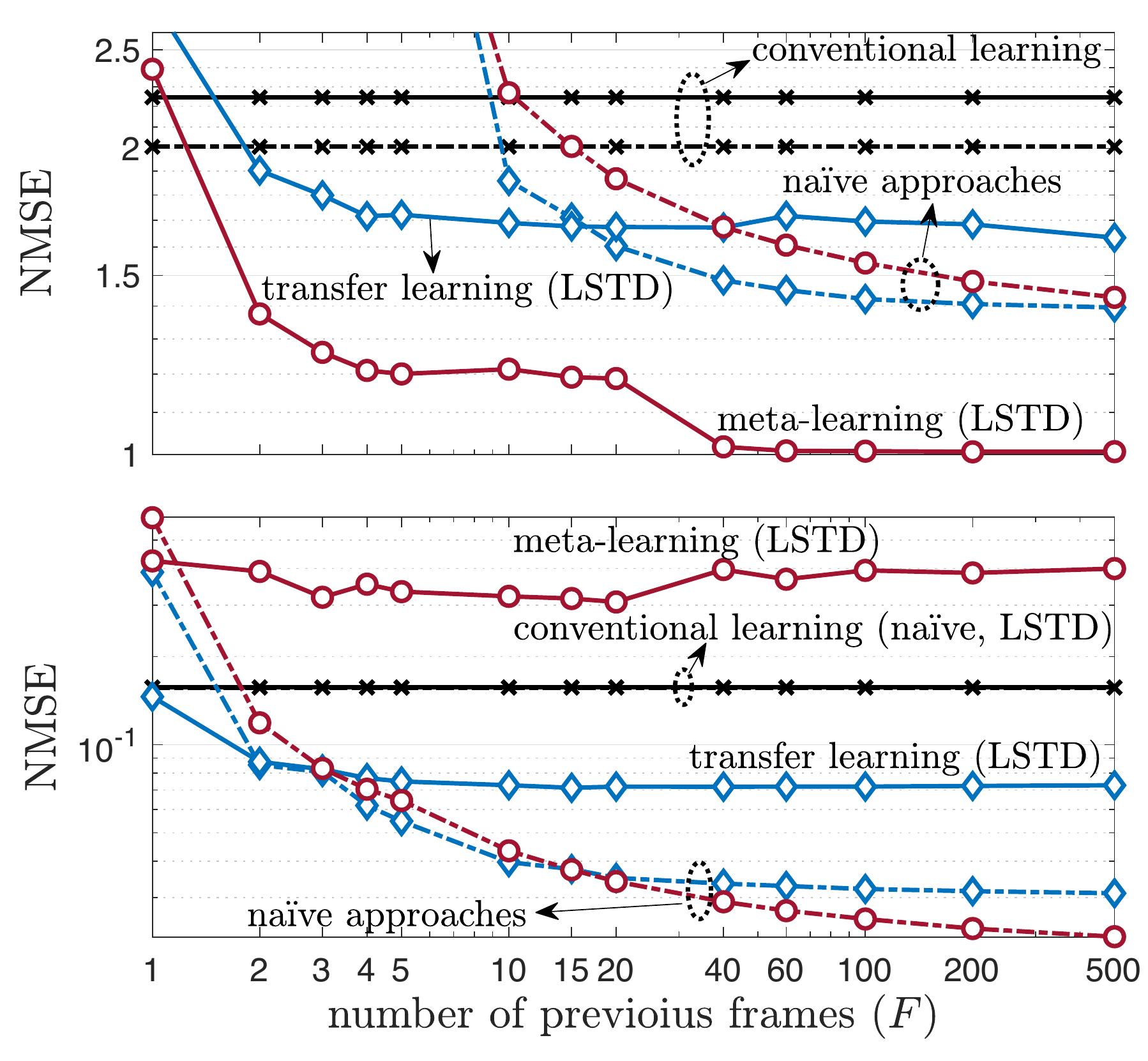}
    \caption{Multi-antenna frequency-selective channel prediction performance as a function of the number of available previous frames $F$ under $19$-clustered, two taps ($W=2$), multi-antenna ($N_T=4, N_R=2$) 3GPP SCM channel model for a fast-varying environment (top) and slow-varying environment (bottom) with number of training samples $L^\text{new}=1$.}
    \label{fig:vec_per_F}
\end{figure}

We now consider the prediction performance for multi-antenna frequency-selective channels as a function of the number of training samples $L^\text{new}$ in Fig.~\ref{fig:vec_per_L_fast} and Fig.~\ref{fig:vec_per_L_slow}, as well as versus the number of frames $F$ in Fig.~\ref{fig:vec_per_F}. For meta-learning, we set $L^\text{tr} = L^\text{new}$ in order to avoid discrepancies between
meta-training and meta-testing \cite{park2020learning}. Fig.~\ref{fig:vec_per_L_fast} and Fig.~\ref{fig:vec_per_L_slow} shows that meta-learning and transfer learning, which utilize $F = 500$ previous frames, can significantly outperform conventional learning in terms of number of
required pilots $L^\text{new}$. This key observation motivates the use of transfer and meta-learning in the presence of limited training data. Furthermore, confirming the analysis in Sec.~\ref{sec:scalar_meta} and Sec.~\ref{sec:meta_lstd}, meta-learning can outperform all other schemes as long as one selects a na\"ive parametrization for slowly-varying environments, and the LSTD parametrization for fast-varying environments. For sufficiently large $L^\text{new}$, transfer learning can, however, improve over meta-learning on fast-varying environments as seen in Fig.~\ref{fig:vec_per_L_fast}. This stems from the split of training and testing set applied by meta-learning, which can lead to a performance loss as $L^\text{new}$ increases.

Lastly, we investigate the effect of number of previous frames $F$ for transfer and meta-learning. As a general result, as demonstrated by Fig.~\ref{fig:vec_per_F}, an increase in the number $F$ of previous frames results in better performance
for both transfer and meta-learning. Furthermore, in a slow-varying environment with a small value of $F$, transfer learning can outperform meta-learning due to the limited need for adaptation, while meta-learning with the correctly select type of parametrization, outperforms transfer learning otherwise. 


\section{Conclusion}
\label{sec:conclusion}
In this paper, we have introduced data-driven channel prediction strategies for multi-antenna frequency-selective channels that aim at reducing the number of pilots by integrating transfer and meta-learning with a novel parametrization of linear predictors. The methods leverage the underlying structure of the wireless channels, which can be expressed in terms of a long-short-term decomposition (LSTD) into long-term space-time features and fading amplitudes. To enable transfer and meta-learning under an LSTD-based model, we have proposed an optimization strategy based on equilibrium propagation (EP) and alternating least squares (ALS). Numerical experiments have shown that the proposed LSTD-based transfer and meta-learning methods far outperform conventional prediction methods, especially in the few-pilots regime. For instance, under a standard 3GPP SCM channel model, assuming four transmit antennas and two receive antennas, using only one pilot meta-learning with LSTD can reduce the normalized prediction MSE by 3 dB as compared to standard learning techniques.
Future work may consider the joint use of deep neural networks, in lieu of linear prediction filters, although related results for multi-antenna frequency-flat channels have not reported any significant advantage to date \cite{jiang2019comparison, jiang2020long, kibugi2021machine, kim2020massive}.
\appendices
\section{Derivation for Long-Short-Term-Decomposition (LSTD)-based Predictor $V_f^{(K)}$}
\label{app:vf_rr}
Since $k$-th amplitudes $\hat{d}_{l-N+1,f}^k,\ldots,\hat{d}_{l+\delta,f}^k$ are all scalar values, we can write $\text{vec}([\hat{d}_{l,f}^k, \ldots, \hat{d}_{l-N+1,f}^k] ) = [\hat{d}_{l,f}^k, \ldots, \hat{d}_{l-N+1,f}^k]^\top$ and 
$\hat{d}_{l+\delta,f}^k = (\hat{d}_{l+\delta,f}^k)^\top$. Using these equalities, we can plug-in \eqref{eq:pred_amplitude} and \eqref{eq:past_amplitude} to \eqref{eq:pred_channel_est_vec}, to get the expression of the predicted channel $\hat{h}_{l+\delta,f}$ as
\begin{align}
    \hat{h}_{l+\delta,f} &= \sum_{k=1}^K \hat{b}_f^k (\hat{b}_f^k)^\dagger H_{l,f}^N ((v_f^k)^\dagger)^\top  \nonumber\\&= \underbrace{\sum_{k=1}^K \left((v_{f}^k)^\dagger \otimes (\hat{b}_f^k (\hat{b}_f^k)^\dagger)\right)}_{(V_f^{(K)})^\dagger \text{ from \eqref{eq:predicted_channel_reduced_rank} }} \text{vec}(H_{l,f}^N).
\end{align}
from which we can easily obtain LSTD-based predictor matrix  $V_f^{(K)} = \sum_{k=1}^K v_{f}^k \otimes (\hat{b}_f^k (\hat{b}_f^k)^\dagger)$.

\section{Details on Conventional Learning for LSTD-based prediction}
\label{app:lstd_conven}
Recalling from \eqref{eq:training_dataset} that $X_f^\text{tr} = [\text{vec}(H_{1,f}^N)^\dagger,\ldots,\text{vec}(H_{L^\text{tr},f}^N)^\dagger]^\top$, we can rewrite one part of ALS \eqref{eq:als_w} in the form of standard ridge regression formula as
\begin{align}
    v_f^{k} &\leftarrow \hspace{-0.8cm}\argmin_{\substack{{v}_f^{k}\\ (V_f^{(K)})^k = {v}_{f}^k \otimes (\hat{b}_f^k (\hat{b}_f^k))^\dagger}} \hspace{-0.8cm}\Big\{ || X_f^\text{tr} (V_f^{(K)})^k - (Y_f^\text{tr})^k ||_F^2 + \lambda_2 || {v}_f^k - \bar{v}^k ||^2 \Big\} \nonumber\\
    &= \argmin_{v_f^k} \Big\{ \sum_{i=1}^{L^\text{tr}} \norm{\hat{b}_{f}^k (\hat{b}_f^{k})^\dagger  H_{i,f}^N ((v_f^k)^\dagger)^\top - (y_{i,f}^\text{tr})^k }^2 + \lambda_2  \norm{v_f^k - \bar{v}^k}^2 \Big\}
    \label{eq:als_v_simple}
\end{align}
with $(y_{i,f}^\text{tr})^k$ being the Hermitian transposition of the $i$-th row of the $k$-th residual target matrix $(Y_f^\text{tr})^k$ defined in \eqref{eq:conven_resi_target}, 
which can be solved in closed-form similar to \eqref{eq:general_ridge_sol_scalar} as
\begin{align}
    ((v_f^k)^\dagger)^\top &=\left(\sum_{i=1}^L (H_{i,f}^N)^\dagger \hat{b}_{f}^k (\hat{b}_{f}^k)^\dagger H_{i,f}^N + \lambda_2 I \right)^{-1} \left( \sum_{i=1}^L (H_{i,f}^N)^\dagger \hat{b}_{f}^k (\hat{b}_{f}^k)^\dagger (y^\text{tr}_{i,f})^k+ \lambda_2 \bar{v}^k \right).
\end{align}

Similarly, the other part of ALS \eqref{eq:als_b} can be rewritten as 
\begin{align}
    \hat{b}_f^k  &\leftarrow \argmin_{\hat{b}_f^k} \Big\{ \sum_{i=1}^{L^\text{tr}} \norm{\hat{b}_{f}^k (\hat{b}_f^{k})^\dagger  H_{i,f}^N ((v_f^k)^\dagger)^\top - (y_{i,f}^\text{tr})^k }^2  - \lambda_1  \text{tr}\left((\hat{b}_f^k)^\dagger (\bar{b}^k (\bar{b}^k)^\dagger) \hat{b}_f^k\right) \Big\} \nonumber\\
    &\quad\quad\quad\quad\quad\quad\quad\quad\quad\quad\quad\quad\quad\quad\quad\quad\quad\quad\quad\quad\quad\quad\quad\quad\text{subject to } b_k^\dagger b_k = 1  \nonumber\\
    &=\argmin_{\hat{b}_f^k}  \Big\{ (\hat{b}_f^k)^\dagger ((\check{X}_{f}^\text{tr})^k)^\dagger (\check{X}_{f}^\text{tr})^k - ((\check{Y}^\text{tr}_{f})^k)^\dagger (\check{X}_{f}^\text{tr})^k - ((\check{X}_{f}^\text{tr})^k)^\dagger (\check{Y}^\text{tr}_{f})^k) \hat{b}_f^k    \Big \}  \label{eq:ALS-b}\nonumber\\ 
    &\quad\quad\quad\quad\quad\quad\quad\quad\quad\quad\quad\quad\quad\quad\quad\quad\quad\quad\quad\quad\quad\quad\quad\text{ subject to } b_k^\dagger b_k = 1,
\end{align}
for which the solution of \eqref{eq:ALS-b} can be obtained by taking the eigenvector of $((\check{X}_{f}^\text{tr})^k)^\dagger (\check{X}_{f}^\text{tr})^k - ((\check{Y}^\text{tr}_{f})^k)^\dagger (\check{X}_{f}^\text{tr})^k - ((\check{X}_{f}^\text{tr})^k)^\dagger (\check{Y}^\text{tr}_{f})^k) $ that corresponds to the smallest eigenvalue, with the matrices $\check{X}_k$ and $\check{Y}_{k,f}$ defined as
\begin{align}
\check{X}_k = \begin{pmatrix}
    (\check{X}_{f}^\text{tr})^k \\
    \sqrt{\lambda_1} \bar{b}_k^\dagger
\end{pmatrix},
(\check{Y}^\text{tr}_{f})^k = \begin{pmatrix}
    (Y_{f}^\text{tr})^k  \\
    \sqrt{\lambda_1} \bar{b}_k^\dagger
\end{pmatrix},
\end{align}
where we denote $(\check{X}_{f}^\text{tr})^k = [(\check{x}_{1,f}^k)^\dagger,\ldots,(\check{x}_{L^\text{tr},f}^k)^\dagger]^\top$ given $\check{x}_{i,f}^k = H_{i,f}^N ((v_{f}^k)^\dagger)^\top$.

\section{Details on Transfer Learning for LSTD-based prediction}
\label{app:lstd_transfer}
Solution of \eqref{eq:general_ridge_vec_rr_seq_trans} can be directly obtained with the tools in Appendix~\ref{app:lstd_conven} given $\lambda_1, \lambda_2 = 0$ and by substituting $X_f^\text{tr}$ and $Y_f^\text{tr}$ with $[X_1, \ldots, X_F]^\top$ and $[Y_1, \ldots, Y_F]^\top$, respectively. 

\section{Details on Meta-Learning for LSTD-based prediction}
\label{app:lstd_meta}
Before deriving \eqref{eq:ep_grad_b} and \eqref{eq:ep_grad_v}, for ease of representation, let us define \emph{inner loss function} $\mathcal{L}_f^\text{inner}$, \emph{outer loss function} $\mathcal{L}_f^\text{outer}$, and \emph{total loss function} $\mathcal{L}_f^\text{total}$ as
\begin{align}
    \mathcal{L}_f^\text{inner}(\hat{b}_f^k, v_f^k | \bar{b}^k, \bar{v}^k) =  \norm{ \frac{X_f^\text{tr} (V_f^{(K)})^k}{(\hat{b}_f^k)^\dagger \hat{b}_f^k} - (Y_f^\text{tr})^k }_F^2  - \lambda_1 \frac{  (\hat{b}_f^k)^\dagger (\bar{b}^k (\bar{b}^k)^\dagger) \hat{b}_f^k }{(\hat{b}_f^k)^\dagger \hat{b}_f^k}+ \lambda_2 || {v}_f^k - \bar{v}^k ||^2 ;
    \label{eq:inner_app}
\end{align}
\begin{align}
    \mathcal{L}_f^\text{outer}(\hat{b}_f^k, v_f^k) =  \norm{ \frac{X_f^\text{te} (V_f^{(K)})^k}{(\hat{b}_f^k)^\dagger \hat{b}_f^k} - (Y_f^\text{te})^k }_F^2 ;
    \label{eq:outer_app}
\end{align}
and 
\begin{align}
    \mathcal{L}_f^\text{total}(\hat{b}_f^k, v_f^k|\bar{b}^k, \bar{v}^k, \alpha)=  \mathcal{L}_f^\text{inner}(\hat{b}_f^k, v_f^k | \bar{b}^k, \bar{v}^k) &+ \alpha \mathcal{L}_f^\text{outer}(\hat{b}_f^k, v_f^k) ,
    \label{eq:total_app}
\end{align}
respectively. Since \eqref{eq:inner_app} is scale-invariant to $(\hat{b}_f^k)^\dagger \hat{b}_f^k$ (recall that $(V_f^{(K)})^k = v_f^k \otimes (\hat{b}_f^k (\hat{b}_f^k)^\dagger)$), this can be considered as an unconstrained version of \eqref{eq:general_ridge_vec_rr_seq}, i.e., $(\hat{b}_f^{k,*}, v_f^{k,*})$ in \eqref{eq:general_ridge_vec_rr_seq} minimizes \eqref{eq:inner_app}. Analogously, \eqref{eq:total_app} can be considered as an unconstrained expression of \eqref{eq:general_ridge_vec_rr_seq_meta_alpha} as $(\hat{b}_f^{k,\alpha}, v_f^{k,\alpha})$ in \eqref{eq:general_ridge_vec_rr_seq_meta_alpha}  minimizes \eqref{eq:total_app}. 

Lastly, it is worth noting that 
\begin{align}
    \sum_{f=1}^F \frac{\partial}{\partial \alpha}\mathcal{L}_f^\text{total} (\hat{b}_f^{k,*}, v_f^{k,*}|\bar{b}^k, \bar{v}^k, 0)
    \label{eq:obj_meta_app}
\end{align}
is equivalent to the objective function of meta-learning in \eqref{eq:general_ridge_vec_rr_seq_meta}.

The following proof for deriving \eqref{eq:ep_grad_b} and \eqref{eq:ep_grad_v} basically follows \cite{scellier2017equilibrium}.
Assuming that the conditions of implicit function theorem \cite{lorraine2020optimizing} are satisfied with respect to $(\hat{b}_f^{k,\alpha}, v_f^{k,\alpha})$, we can write cross-derivatives of $\mathcal{L}_f^{\text{total}}(\hat{b}_f^{k,\alpha}, v_f^{k,\alpha}|\bar{b}^k, \bar{v}^k, \alpha)$ as 
\begin{align}
    \frac{d}{d \bar{b}^k} \frac{\partial \mathcal{L}_f^\text{total}}{\partial \alpha} (\hat{b}_f^{k,\alpha}, v_f^{k,\alpha} | \bar{b}^k, \bar{v}^k, \alpha) = \frac{d}{d \alpha} \frac{\partial \mathcal{L}_f^\text{total}}{\partial \bar{b}^k} (\hat{b}_f^{k,\alpha},v_f^{k,\alpha} | \bar{b}^k, \bar{v}^k, \alpha)
    \label{eq:app_ep_b}
\end{align}
and
\begin{align}
    \frac{d}{d \bar{v}^k} \frac{\partial \mathcal{L}_f^\text{total}}{\partial \alpha} (\hat{b}_f^{k,\alpha}, v_f^{k,\alpha} | \bar{b}^k, \bar{v}^k, \alpha) = \frac{d}{d \alpha} \frac{\partial \mathcal{L}_f^\text{total}}{\partial \bar{v}^k} (\hat{b}_f^{k,\alpha},v_f^{k,\alpha} | \bar{b}^k, \bar{v}^k, \alpha),
    \label{eq:app_ep_v}
\end{align}
from the property of stationary point  $(\hat{b}_f^{k,\alpha}, v_f^{k,\alpha})$ in \eqref{eq:general_ridge_vec_rr_seq_meta_alpha}, i.e.,
\begin{align}
      \frac{\partial \mathcal{L}_f^\text{total}}{\partial \hat{b}_f^k} (\hat{b}_f^{k,\alpha}, v_f^{k,\alpha} | \bar{b}^k, \bar{v}^k, \alpha) = \frac{\partial \mathcal{L}_f^\text{total}}{\partial v_f^k} (\hat{b}_f^{k,\alpha}, v_f^{k,\alpha} | \bar{b}^k, \bar{v}^k, \alpha) = 0.
\end{align}
Recalling \eqref{eq:obj_meta_app}, the left-hand side of \eqref{eq:app_ep_b} with $\alpha=0$ consists summand of left-hand side of \eqref{eq:ep_grad_b}, and similarly, left-hand side of \eqref{eq:app_ep_v} with $\alpha=0$ composes left-hand side of \eqref{eq:ep_grad_v}. This implies that, if we can compute \eqref{eq:app_ep_b} and \eqref{eq:app_ep_v}, we can get the desired gradients for meta-learning. To this end, we first compute the partial derivative terms of \eqref{eq:app_ep_b} and \eqref{eq:app_ep_v} from \eqref{eq:inner_app}--\eqref{eq:total_app} as
\begin{align}
    \frac{\partial \mathcal{L}_f^\text{total}}{\partial \bar{b}^k} (\hat{b}_f^{k,\alpha},v_f^{k,\alpha} | \bar{b}^k, \bar{v}^k, \alpha) = 2\lambda_1(\bar{b}^k - \hat{b}_f^{k,\alpha} (\hat{b}_f^{k,\alpha})^\dagger \bar{b}^k)
    \label{eq:app_ep_bbar}
\end{align}
for \eqref{eq:app_ep_b} and
\begin{align}
    \frac{\partial \mathcal{L}_f^\text{total}}{\partial \bar{v}^k} (\hat{b}_f^{k,\alpha},v_f^{k,\alpha} | \bar{b}^k, \bar{v}^k, \alpha) = 2\lambda_2 ( \bar{v}^k - v_f^{k,\alpha}),
    \label{eq:app_ep_vbar}
\end{align}
for \eqref{eq:app_ep_v}. Next step is to use finite differentiation method \cite{scellier2017equilibrium} to obtain the desired gradients for meta-learning as
\begin{align}
    \nabla_{\bar{b}^k}&\sum_{f=1}^F \norm{  X_f^\text{te} (V_f^{(K)})^{k,*} - (Y_{f}^\text{te})^k}_F^2 \nonumber\\&= \lim_{\alpha \rightarrow 0} \sum_{f=1}^F \frac{1}{\alpha} \Big( \frac{\partial \mathcal{L}_f^\text{total}}{\partial \bar{b}^k} (\hat{b}_f^{k,\alpha},v_f^{k,\alpha} | \bar{b}^k, \bar{v}^k, \alpha) -  \frac{\partial \mathcal{L}_f^\text{total}}{\partial \bar{b}^k} (\hat{b}_f^{k,*},v_f^{k,*} | \bar{b}^k, \bar{v}^k, 0) \Big) 
    \nonumber\\&=  \lim_{\alpha \rightarrow 0} \frac{2\lambda_1}{\alpha} \sum_{f=1}^F \left( \hat{b}_f^{k,*} (\hat{b}_{f}^{k,*})^\dagger - \hat{b}_{f}^{k,\alpha} (\hat{b}_{f}^{k,\alpha})^\dagger \right)\bar{b}^k
    \label{eq:ep_grad_b_app}
\end{align}
and 
\begin{align}
    \nabla_{\bar{v}^k}&\sum_{f=1}^F \norm{  X_f^\text{te} (V_f^{(K)})^{k,*} - (Y_{f}^\text{te})^k}_F^2  \nonumber\\&= \lim_{\alpha \rightarrow 0} \sum_{f=1}^F \frac{1}{\alpha} \Big( \frac{\partial \mathcal{L}_f^\text{total}}{\partial \bar{v}^k} (\hat{b}_f^{k,\alpha},v_f^{k,\alpha} | \bar{b}^k, \bar{v}^k, \alpha) -  \frac{\partial \mathcal{L}_f^\text{total}}{\partial \bar{v}^k} (\hat{b}_f^{k,*},v_f^{k,*} | \bar{b}^k, \bar{v}^k, 0) \Big) 
    \nonumber\\&=  \lim_{\alpha \rightarrow 0} \frac{2\lambda_2}{\alpha} \sum_{f=1}^F(v^{k,*}_f - v^{k,\alpha}_f),
    \label{eq:ep_grad_v_app}
\end{align}
which concludes the derivation of \eqref{eq:ep_grad_b} and \eqref{eq:ep_grad_v}. Generalized proof along with useful characteristics of EP can be found in \cite{scellier2017equilibrium}.

\section{Details on the Antenna Configuration in Sec.~\ref{sec:exp_per_antenna}}
\label{app:per_antenna_exp}
Following the table contains the specification of the antenna configurations in Sec.~\ref{sec:exp_per_antenna}. We denote  $(N_R^\text{hor}, N_R^\text{ver},N_R^\text{pol}, N_T^\text{hor}, N_T^\text{ver}, N_T^\text{pol})$ by the pair of number of horizontal receive antennas $N_R^\text{hor}$, number of vertical receive antennas $N_R^\text{ver}$,  number of polarizations of receive antennas $N_R^\text{pol}$, number of horizontal transmit antennas $N_T^\text{hor}$, number of vertical transmit antennas $N_T^\text{ver}$, and number of polarizations of transmit antennas $N_T^\text{pol}$. Note that $N_RN_T = N_R^\text{hor} N_R^\text{ver}N_R^\text{pol} N_T^\text{hor} N_T^\text{ver} N_T^\text{pol}$.

\begin{table}[h!]
\caption{Antenna Configurations for Sec.~\ref{sec:exp_per_antenna}}
\label{table:exp_settings_antenna}
\centering
\begin{center}
\begin{tabular}{||c | c||} 
 \hline
 number of & antenna configuration\\total antennas ($N_R N_T$)   &$(N_R^\text{hor}, N_R^\text{ver},N_R^\text{pol}, N_T^\text{hor}, N_T^\text{ver}, N_T^\text{pol})$ \\ [0.5ex] 
 \hline\hline
 1 & $(1,1,1,1,1,1)$ \\ 
 \hline
 2 & $(1,1,1,2,1,1)$ \\
 \hline
 4 & $(1,1,1,2,2,1)$ \\
 \hline
 8 & $(2,1,1,2,2,1)$ \\
 \hline
 16 & $(2,1,1,2,2,2)$  \\
 \hline
  32 & $(2,1,1,4,2,2)$  \\
 \hline
  64 & $(2,1,1,4,4,2)$  \\
 \hline
  128 & $(2,2,1,4,4,2)$ \\
 \hline
\end{tabular}
\end{center}
\end{table}


\bibliographystyle{IEEEtran}
\bibliography{ref}
\end{document}